\documentclass[useAMS,usenatbib,aas_macros]{mn2e}
\usepackage{aas_macros}

\usepackage{color}
\newcommand{\adb}[1]{#1}

\newcommand{\equ}[1]{eq.~(\ref{eq:#1})}

\newcommand{\equnp}[1]{eq.~\ref{eq:#1}}
\newcommand{\se}[1]{\S\ref{sec:#1}}
\newcommand{\fig}[1]{Fig.~\ref{fig:#1}}
\newcommand{\figs}[1]{Figs.~\ref{fig:#1}}
\newcommand{\Fig}[1]{Figure~\ref{fig:#1}}
\newcommand{\Figs}[1]{Figures~\ref{fig:#1}}
\newcommand{\tab}[1]{Table~\ref{tab:#1}}
\newcommand{\be}{\begin{equation}}
\newcommand{\ee}{\end{equation}}
\newcommand{\bea}{\begin{eqnarray}}
\newcommand{\eea}{\end{eqnarray}}
\def\ra{\rangle}
\def\la{\langle}
\def\med{{\rm median}}

\newcommand{\msun}{M_\odot}

\newcommand{\sy}{\,M_\odot\, {\rm yr}^{-1}}

\newcommand{\ifm}[1]{\relax\ifmmode#1\else$\mathsurround=0pt #1$\fi}
\newcommand{\kms}{\ifmmode\,{\rm km}\,{\rm s}^{-1}\else km$\,$s$^{-1}$\fi}
\newcommand{\hmpc}{\,\ifm{h^{-1}}{\rm Mpc}}

\newcommand{\Mpc}{\,{\rm Mpc}}

\newcommand{\kpc}{\,{\rm kpc}}
\newcommand{\pc}{\,{\rm pc}}
\newcommand{\Gyr}{\,{\rm Gyr}}

\newcommand{\Myr}{\,{\rm Myr}}

\newcommand{\cmc}{\,{\rm cm}^{-3}}

\newcommand{\ltsima}{$\; \buildrel < \over \sim \;$}
\newcommand{\lsim}{\lower.5ex\hbox{\ltsima}}
\newcommand{\gtsima}{$\; \buildrel > \over \sim \;$}
\newcommand{\gsim}{\lower.5ex\hbox{\gtsima}}
\newcommand{\prop}{\propto}

\newcommand{\rar}{\rightarrow}

\def\omm{\Omega_{\rm m}}
\def\omb{\Omega_{\rm b}}
\def\oml{\Omega_{\Lambda}}

\def\Mv{M_{\rm v}}

\def\Rv{R_{\rm v}}
\def\Vv{V_{\rm v}}
\def\Tv{T_{\rm v}}
\def\Mg{M_{\rm g}}
\def\Ms{M_{\rm star}}
\def\Mps{M_{\rm ps}}

\def\Mdot{\dot{M}}
\def\Mt{M_{\rm tot}}
\def\Mb{M_{\rm b}}
\def\Md{M_{\rm d}}
\def\Rd{R_{\rm d}}
\def\Vd{V_{\rm d}}
\def\fg{f_{\rm g}}

\def\Mc{M_{\rm c}}

\def\tv{t_{\rm v}}
\def\td{t_{\rm d}}
\def\tff{t_{\rm ff}}

\def\epsf{\epsilon_{\rm sf}}
\def\fsf{f_{\rm sf}}
\def\tmig{t_{\rm mig}}
\def\tlambda{\widetilde{\lambda}}
\def\tnu{\upsilon}

\def\fb{f_{\rm b}}

\def\ga{s}

\title[Galaxy-Formation Toy Models]
{Toy Models for Galaxy Formation versus Simulations}

\author[Dekel et al.]
{
A. Dekel$^{1}$,
A. Zolotov$^{1}$,
D. Tweed$^{1}$,
M. Cacciato$^{1,2}$,
D. Ceverino$^{1,3}$,
J.R. Primack$^{4}$
\\
\\
$^1$Racah Institute of Physics, The Hebrew University, Jerusalem 91904 Israel;
(avishai.dekel@mail.huji.ac.il)\\
$^2$Leiden Observatory, Leiden University, 
       Niels Bohrweg 2, NL-2333 CA Leiden, The Netherlands\\
$^3$Departamento de Fisica Teorica, Universidad Autonoma de Madrid, 
Madrid E-28049, Spain\\
$^4$Department of Physics, University of California, Santa Cruz, CA, 95064, USA
}

\begin{document}

\large

\pagerange{\pageref{firstpage}--\pageref{lastpage}} \pubyear{2002}

\maketitle

\label{firstpage}

\begin{abstract}
We describe simple useful toy models for key processes of galaxy formation in 
its most active phase, at $z>1$, and test the approximate expressions against 
\adb{the typical behaviour in a suite of}
high-resolution hydro-cosmological simulations of massive galaxies at $z=4-1$.
We address in particular the evolution of (a) the total mass inflow rate from 
the cosmic web into galactic haloes based on the EPS approximation, (b) the 
penetration of baryonic streams into the inner galaxy, (c) the disc size, (d) 
the implied steady-state gas content and star-formation rate (SFR) in the 
galaxy subject to mass conservation and a universal star-formation law, (e) the
inflow rate within the disc to a central bulge and black hole as derived using 
energy conservation and self-regulated $Q\sim 1$ violent disc instability (VDI),
and (f) the implied steady state in the disc and bulge. The toy models provide 
useful approximations for the behaviour of the simulated galaxies. We find that
(a) the inflow rate is proportional to mass and to $(1+z)^{5/2}$, (b) the 
penetration to the inner halo is $\sim 50\%$ at $z=4-2$, (c) the disc radius is
$\sim 5\%$ of the virial radius, (d) the galaxies reach a steady state with the
SFR following the accretion rate into the galaxy, (e) there is an intense gas 
inflow through the disc, comparable to the SFR, following the predictions of 
VDI, and (f) the galaxies approach a steady state with the bulge mass 
comparable to the disc mass, where the draining of gas by SFR, outflows and 
disc inflows is replenished by fresh accretion. Given the agreement with
simulations, these toy models are useful for understanding the complex 
phenomena in simple terms and for back-of-the-envelope predictions. 
\end{abstract}


\section{Introduction}
\label{sec:intro}

A scientific understanding of a physical phenomenon is often materialized
through a simple, idealized, analytic model, which we call a toy model.
A toy model is derived from first
principles in an attempt to capture the key physical elements of the process 
and to provide a useful approximation to the complex behaviour 
seen in numerical 
simulations or in observations.
The simple scaling relations from
such a toy model allow back-of-the-envelope calculations that involve
several elements or a sequence of events, which may help 
to develop an understanding for a whole scenario, and hopefully lead
to explanations of observed phenomena or to new theoretical predictions.

\smallskip 
\adb{
A different tool in the study of galaxy formation is semi-analytic modeling 
\citep[SAM, e.g.,][]{somerville08,guo11,benson12}.
In SAMs, the many complex physical 
processes associated with gas, stars, black holes and radiation are 
modeled in parallel by recipes with free parameters, 
incorporated in retrospect in dark-matter halo merger trees that were obtained 
from dissipationless cosmological N-body simulations or the Extended
Press-Schechter approximation \citep[EPS,][]{bond91}. 
The toy models addressed here are much simpler than the sophisticated SAMs. 
The toy models are typically based on simple analytic arguments using crude 
approximations, and they tend to focus on one process at a time. These toy 
models could serve as a basis for physical recipes to be incorporated in SAMs.  
SAM recipes have been confronted with the more
elaborate hydro-cosmological simulations, where gas processes are explicitly 
simulated
\citep[e.g.][]{benson01,cattaneo07,neistein12,hirschmann12},
but this is commonly done in the context of a full SAM.
Here we compare each toy model on its own to the corresponding feature of the
simulation results.
}

\smallskip
We 
\adb{address}
toy models for the formation and evolution of galaxies in their
most active phase, at redshifts $z>1$, where the Einstein-deSitter (EdS)
cosmological model serves as a useful approximation.
\adb{We focus on central galaxies in their haloes, the main progenitors
of central galaxies at $z \sim 1$.}
In particular, we address several key processes as follows. 
(a) We verify the inflow rate of dark matter and baryons from the cosmic web 
into 
\adb{distinct}
haloes based on the EPS approximation. 
(b) We evaluate the penetration of baryonic streams through the hot gas in the
haloes into the inner disc galaxies.
(c) We crudely estimate the disc radius as a function of the virial radius
compared to the standard model where angular momentum is conserved.
(d) We study the implied gas content and star-formation rate (SFR) in the 
galaxy subject to mass conservation (the ``bathtub" model)
and a universal SFR law.  
(e) We address the inflow rate within the disc to a central bulge as 
derived using energy conservation and Toomre Violent Disc Instability (VDI).
(f) We study the implied steady state in the disc as it is fed by cold streams
and is drained by star formation, outflows, and inflow to the bulge.

\smallskip
To test the validity of the toy models,
we compare their predictions with the results obtained from a suite of 27
Adaptive Mesh Refinement (AMR) hydro-cosmological simulations 
that zoom in on individual galaxies
with a resolution of $\sim 50\pc$.
These are state-of-the-art simulations that resolve the key processes 
addressed. 
The feedback in these simulations is limited to 
the thermal effects of stellar winds and supernovae,
not addressing radiative feedback and AGN feedback,
thus not allowing an immediate study of the effects 
of very strong outflows and the associated severe suppression of SFR.

\smallskip
The outline of this paper is as follows.
In \se{sims} we describe the simulations.
In \se{accretion} we address the mass inflow rate into haloes.
In \se{penetration} we study the penetration of cold streams into the 
central galaxy at the inner halo.
In \se{spin} we refer to the disc size as derived from conservation of 
specific angular momentum.
In \se{SS} we investigate the steady-state of the gas and stellar content
in which the SFR follows the accretion rate.
In \se{inflow} we address the VDI-driven inflow within the disc into the 
bulge.
In \se{SS_disc} we refer to the resultant steady state in the disc.
In \se{conc} we summarize our conclusions.

\section{Simulations}
\label{sec:sims}

We use zoom-in hydro cosmological simulations of 27 moderately massive
galaxies with an AMR maximum resolution $35-70\pc$, all evolved 
\adb{from high redshift}
to $z=2$, many to $z \sim 1.5$ and some reaching $z=1$, 
as marked by $a_{\rm fin}$ in \tab{sim_table}.
They utilize the Adaptive Refinement Tree (ART) code
\citep{krav97,ceverino09}, 
which accurately follows 
the evolution of a gravitating N-body system and the Eulerian gas dynamics 
using an adaptive mesh.
Beyond gravity and hydrodynamics, the code incorporates at the subgrid level
many of the physical processes relevant for galaxy formation.
They include gas cooling by atomic hydrogen and helium as well as by
metals and molecular hydrogen, photoionization heating by the UV background 
with partial self-shielding, 
star formation, stellar mass loss, metal enrichment of the ISM,
and feedback from stellar winds and supernovae, implemented as local injection
of thermal energy 
\citep{ceverino09,cdb10,ceverino12}.

\begin{table*}
\centering
  \begin{tabular}{cccccccc}
      \hline
  Galaxy &$\Rv$&$\Mv$&$\Ms$&$\Mg$& &$a_{\rm fin}$&$\Mv(a_{\rm fin})$\\
         &kpc&$10^{12}\msun$&$10^{11}\msun$&$10^{11}\msun$& & &$10^{12}\msun$\\
      \hline
      MW01&102&0.81&0.72&0.57& &0.42&1.39\\
      MW02&105&0.89&2.56&1.12& &0.34&0.94\\
      MW03&099&0.73&0.60&0.51& &0.42&1.34\\
      MW04&123&1.42&1.41&0.89& &0.38&1.93\\
      MW07&073&0.30&0.30&0.22& &0.40&0.49\\
      MW08&071&0.28&0.28&0.15& &0.45&0.59\\
      MW09&059&0.16&0.19&0.08& &0.50&0.64\\
      MW10&102&0.82&0.72&0.44& &0.50&1.63\\
      MW11&088&0.53&0.51&0.28& &0.40&0.61\\
      MW12&130&1.70&2.06&1.01& &0.39&3.49\\
      \hline
      VL01&117&1.23&1.54&0.75& &0.37&1.89\\
      VL02&101&0.81&0.89&0.46& &0.50&2.04\\
      VL03&117&1.22&1.44&0.76& &0.33&1.17\\
      VL04&109&1.01&1.33&0.51& &0.42&1.17\\
      VL05&118&1.28&1.29&0.75& &0.41&2.09\\
      VL06&099&0.75&0.94&0.32& &0.50&1.97\\
      VL07&129&1.66&2.15&0.82& &0.34&1.62\\
      VL08&112&1.09&1.35&0.46& &0.46&1.19\\
      VL09&086&0.49&0.61&0.24& &0.34&0.54\\
      VL10&102&0.81&0.95&0.44& &0.50&3.55\\
      VL11&130&1.73&2.02&0.81& &0.50&2.23\\
      VL12&105&0.90&0.96&0.51& &0.50&2.68\\
      \hline
      SFG1&129&1.66&2.10&0.87& &0.38&1.98\\
      SFG4&112&1.09&1.16&0.66& &0.38&1.31\\
      SFG5&123&1.38&1.52&0.78& &0.40&1.73\\
      SFG8&121&1.38&1.70&0.72& &0.35&1.42\\
      SFG9&135&1.89&2.44&1.22& &0.48&5.34\\
      \hline
   \end{tabular}
  \caption{The suite of simulated galaxies.
Quoted at $a=0.33$ ($z=2$) are the virial radius $\Rv$ and within it the 
total mass $\Mv$, stellar mass $\Ms$ and gas mass $\Mg$. 
The snapshots used are from $a=0.2$ ($z=4$) to $a=a_{\rm fin}$ 
($z_{\rm fin}=a_{\rm fin}^{-1}-1 =2-1$),
and the virial mass at $a_{\rm fin}$ is quoted.
\label{tab:sim_table}}
\end{table*}


\begin{figure*}
\vskip 5.5cm
\includegraphics{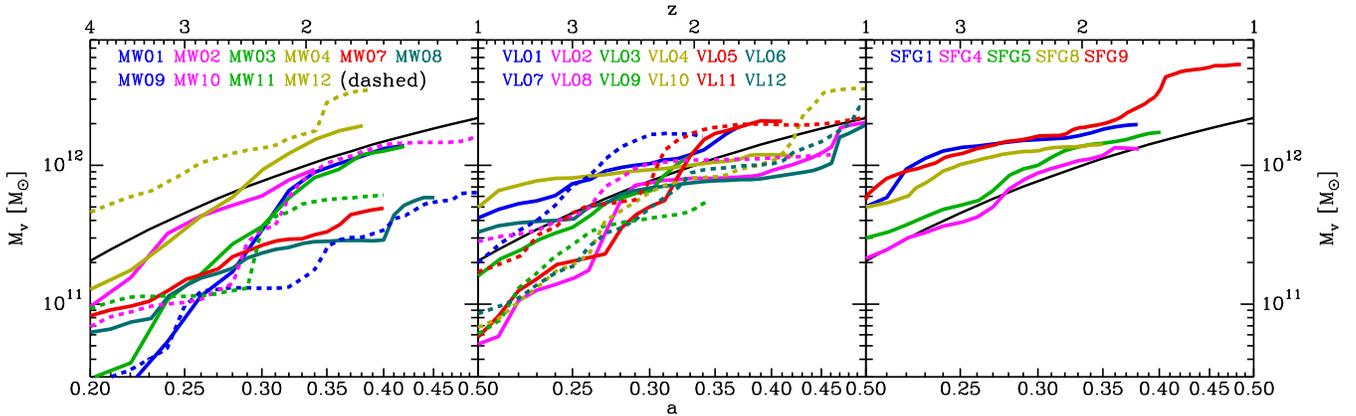}
\caption{
Growth of virial mass $\Mv$ as a function of expansion factor $a=(1+z)^{-1}$
for the individual simulated galaxies.
Each galaxy has been evolved to a final redshift as specified in 
\tab{sim_table}.
The simulations span about a decade in mass, from $\Mv=0.16\times 10^{12}$
to $1.9\times 10^{12}\msun$ at $z=2$.
The solid black curves refer to the toy-model prediction, \equ{mass_toy},
normalized to $\Mv=10^{12}\msun$ at $z=2$.
The overall growth pattern is reproduced by the toy model quite well,
though the individual galaxies can grow in different rates at different
times in their histories.
}
\label{fig:mvir_growth}
\end{figure*}

\subsection{Subgrid Physics}

A few relevant details concerning the subgrid physics are as follows.
Cooling and heating rates are tabulated for a given gas
density, temperature, metallicity and UV background based on the CLOUDY code
\citep{ferland98}, assuming a slab of thickness 1 kpc. A uniform UV background
based on the redshift-dependent \citet{haardt96} model is assumed,
except at gas densities higher than $0.1\cmc$, where a substantially
suppressed UV background is used
($5.9\times 10^{26}{\rm erg}{\rm s}^{-1}{\rm cm}^{-2}{\rm Hz}^{-1}$)
in order to mimic the partial self-shielding of dense gas.
This allows the dense gas to cool down to temperatures of $\sim 300$K.
The assumed equation of state is that of an ideal mono-atomic gas.
Artificial fragmentation on the cell size is prevented by introducing
a pressure floor, which ensures that the Jeans scale is resolved by at least
7 cells \citep[see][]{cdb10}.

\smallskip
Star formation is assumed to occur at 
\adb{cell}
densities above a threshold of $1\cmc$
and at temperatures below $10^4$K. More than 90\% of the stars form at
temperatures well below $10^3$K, and more than half the stars form at 300~K
in cells where the gas density is higher than $10\cmc$.
The code implements a stochastic star-formation model that yields a
star-formation efficiency per free-fall time of 5\%. At the given resolution,
this efficiency roughly mimics 
\adb{by construction}
the empirical Kennicutt-Schmidt law
\citep{schmidt59,kennicutt98}.
\adb{When extracting the resultant approximate SFR, we consider the mass in 
stars born in a given timestep divided by the elapsed time.}
The code incorporates a thermal stellar feedback model, in which the combined
energy from stellar winds and supernova explosions is released as a constant
heating rate over $40\Myr$ following star formation, the typical age of the
lightest star that explodes as a type-II supernova.
The heating rate due to feedback may or may not overcome the cooling
rate, depending on the gas conditions in the star-forming regions
\citep{ds86,ceverino09}. We note that no shutdown of cooling 
\adb{or any other artificial mechanism for boosting the feedback effects}
is implemented in these simulations.
We also include the effect of runaway stars by assigning a velocity kick of
$\sim 10 \kms$ to 30\% of the newly formed stellar particles.
The code also includes the later effects of type-Ia supernova and
stellar mass loss, and it follows the metal enrichment of the ISM.

\subsection{Selected Haloes}

The initial conditions for the high-resolution, zoom-in, hydrodynamical 
simulations that are used in this paper were based on dark-matter haloes that
were drawn from dissipationless N-body simulations at lower resolution
in three large comoving cosmological boxes.
The assumed cosmology is the standard $\Lambda$CDM model with the WMAP5 values
of the cosmological parameters, namely
$\omm=0.27$, $\oml=0.73$, $\omb= 0.045$, $h=0.7$ and $\sigma_8=0.82$
\citep{komatsu09}.
\adb{
Distinct haloes were identified about density peaks, and their virial radius 
and mass, $\Rv$ and $\Mv$, were measured from the radial mass profile 
such that the mean overdensity within the sphere of radius $\Rv$ equals 
$\Delta(a)$ (\se{virial}, \equnp{Delta}). 
We do not explicitly address the evolution of subhaloes in this work, but
rather consider them as part of
the accretion onto the the central galaxy of their host distinct halo.
The haloes for re-simulation were selected to have a given virial mass at $z=1$ 
(or $z=0$ in a few cases), 
as specified in \tab{sim_table}}.
The only other selection criterion was that the haloes show no ongoing major 
merger at $z=1$. This eliminates 
\adb{from the sample} 
less than $10\%$ of the haloes which
tend to be in a dense environment at $z \sim 1$,
and it induces only a minor selection effect at higher redshifts.
The target virial masses at $z=1$ were typically selected to be
$\Mv \sim (1.5-3)\times 10^{12}\msun$, with most of them
intended to end up as $(3-6)\times 10^{12}\msun$ today if left in isolation, 
namely somewhat more massive than the Milky Way. However, the actual mass
range of these haloes at $z=0$ is broader, with some of the haloes 
\adb{destined to merge} 
into more massive haloes hosting groups or clusters.

\subsection{Zoom in}
 
The initial conditions corresponding to each of the selected haloes
were filled with gas and refined to a much higher resolution on an adaptive
mesh within a zoom-in Lagrangian volume that encompasses the mass within 
twice the virial radius at $z =1$, which is roughly a sphere of comoving 
radius $1\Mpc$.
This was embedded in a comoving cosmological box of side 20, 40 or $80\hmpc$.
Each galaxy has been evolved with the full hydro ART and subgrid physics on an
adaptive comoving mesh refined in the dense regions
to cells of minimum size between 35-70 pc in physical units at all times.
This maximum resolution is valid in particular throughout the cold discs
and dense clumps, allowing cooling to $\sim 300$K and gas densities of
$\sim 10^3\,{\rm cm}^{-3}$.
The dark-matter particle mass is $6.7\times 10^5\msun$, 
and the particles representing groups of stars have a minimum mass of 
$10^4\msun$.

\smallskip
The virial properties of all 27 galaxies in our sample 
are listed in \tab{sim_table}.
This includes the virial radius and total virial mass at $z=2$, 
the stellar mass and gas mass within the virial radius at that time, 
the latest time of analysis for each galaxy in terms of the expansion factor,
$a_{\rm fin}$, and the virial mass at that time. 
\adb{
The galaxies MW01-03 were the basis for the study of VDI in \citet{cdb10}.
The galaxies MW04 and SFG1 were added to the study of giant clumps 
in VDI discs by \citet{ceverino12}.
}

\subsection{Analysis}

We start the analysis presented below
at the cosmological time corresponding to expansion
factor $a=0.2$ (redshift $z=4$). At earlier times, the fixed resolution scale
typically corresponds to a larger fraction of the galaxy size, 
which may bias some of the quantities that we wish to study here.
All 27 galaxies reach $a=0.33$ ($z=2$),
17 galaxies reach $a=0.4$ ($z=1.5$),
and only 7 galaxies have been run all the way to $a=0.5$ ($z=1$).
This gradual degradation of the sample after $z=2$ has been taken into 
account in our analysis.

\smallskip
The output of each simulation is provided at output times separated by a 
constant interval in $a$, $\Delta a$, 
which for most galaxies is $\Delta a =0.01$.  
For two galaxies (SFG8-9) 
the timestep is twice as small, $\Delta a =0.005$,
and for four galaxies (MW01-4) the timestep is twice
as large, $\Delta a = 0.02$.
For every galaxy we analyze the data \adb{averaged over} timesteps of 
$\Delta a=0.02$.

\smallskip
\adb{
The quantities addressed, for example, are mass $M$, inflow rate $\Mdot$, and
specific inflow rate $\Mdot/M$.
For comparison with a toy model at a given time step, we stack the data from
all the available galaxies, sometimes after proper scaling motivated by the
model in order to minimize possible mass dependence.
For each quantity, we compute and show the median and the linear average. 
Because of the lognormal nature of the distribution 
(see below, \se{bar_acc}, \fig{acc_Rv_histogram}),
we also compute the average of the log within the 90\% percentiles.
The tails are eliminated here because $\Mdot$ could be negative in a few rare 
cases due, for example, to a fly-by satellite caught while crossing the virial 
radius outwards. The exclusion of the outliers also helps reducing fluctuations 
between time steps. 
The scatter is typically marked in our figures by shaded areas between the 
upper and lower 68\% percentiles and 90\% percentiles.
We also show as error-bars an estimate for the error of the mean,
$\sigma/\sqrt{N}$, where $\sigma$ is the standard
deviation over the $N$ galaxies averaged over at the given time step .
}

\smallskip
The statistical analysis considers all the galaxies of the sample in the time 
range $a=0.2-0.33$ ($z=4-2$), but only the gradually degrading subsample 
between $a=0.33$ ($z=2$) and $a=0.5$ ($z=1$) according to the value of 
$a_{\rm fin}$ in \tab{sim_table}. This translates to larger uncertainties
in the results after $z=2$. 
\adb{
We therefore focus most of our attention on the simulations at $z \geq 2$. 
When addressing detailed disc properties 
(e.g., in \se{inflow_sim} and \se{SS_disc}), 
we limit the main analysis to the range $z=3-2$.
}

\subsection{The Sample of Galaxies}

\Fig{mvir_growth} shows the curves of virial mass growth in time
for the different galaxies listed in \tab{sim_table}.
At $z = 2$,
the sample spans roughly an order of magnitude in halo mass,
from $\Mv=0.16\times 10^{11}$ to $1.9\times 10^{12}\msun$,
with virial radii from $70$ to $135\kpc$.
The simulated haloes typically contain within the virial radius 
a baryon fraction of 0.14, slightly smaller than the cosmic value, 
of which the stellar fraction is typically 0.10.
This suite of galaxies 
spans the mass range of typical observed massive star-forming galaxies at
$z \sim 2$ \citep{forster09}, and otherwise no major selection criteria was
imposed on their properties at $z \sim 2-4$.
As shown in \citet{cdb10}, our first simulated galaxies, MW01-03,
are consistent with the observed scattered scaling relations of $z \sim 2$ 
galaxies, including the relation between
SFR and stellar mass and the Tully-Fisher relation 
\citep[][but see a discussion of deviations below]{forster09,cresci09}.
We can therefore assume that this is approximately a fair sample of galaxies 
in the relevant mass and redshift range, excluding galaxies
that will be in rather dense environments at $z \sim 1$.

\smallskip
The mass growth curves shown in \fig{mvir_growth} are to be visually
compared to the toy model prediction discussed later in \equ{mass_toy},
which is arbitrarily normalized in this figure to $\Mv=10^{12}\msun$ at $z=2$.
The overall growth pattern is similar to the toy model prediction,
though the individual galaxies can grow in different rates at different
times in their histories.
We note that in some cases galaxies tend to keep their rank order 
with respect to the other galaxies in terms of virial mass, but in other cases
galaxies change their rank order drastically during the evolution (e.g., MW03,
left panel, solid green line),
commonly following major mergers in their histories. 
The simulated growth rates are compared to the toy model in more detail
in \figs{total_sMdot_rvir} to \ref{fig:total_Mdot_rvir}. 

\subsection{Limitations of the Current Sample}

These simulations are state-of-the-art in terms of the high-resolution
AMR hydrodynamics and the treatment of key physical processes at the
subgrid level. In particular, they properly trace the cosmological streams
that feed galaxies at high redshift, including mergers and smooth flows,
and they resolve the violent disc instability that governs the high-$z$ disc 
evolution and the bulge formation \citep{cdb10,ceverino12}.
\adb{
When tracing the important small-scale processes involved in galaxy
evolution, AMR codes offer a distinct advantage over SPH codes, 
e.g., because SPH does not accurately model processes such as
shocks and hydrodynamical instabilities 
\citep[e.g.][]{agertz07,scannapieco12,bauer12}.
The AMR codes seem to be comparable in their capabilities to new codes using 
a moving unstructured grid \citep{bauer12}, but the latter are currently
limited to significantly lower resolution, on the order of $\sim 1\kpc$
\citep[e.g.,][]{nelson13},
which is not sufficient for tracing the key processes of disc instabilities 
or the evolution of streams. 
}
 
\smallskip 
However, like other simulations,
\adb{
the current simulations do not yet treat the star formation and feedback
processes with sufficient accuracy.
}
For example, the code assumes a somewhat high SFR efficiency per free-fall
time,
it does not follow in detail the formation of molecules  
and the effect of metallicity on SFR \citep{kd12},
and it does not explicitly include radiative stellar feedback
\citep{murray10,kd10,hopkins12c,dk13}
or AGN feedback \citep{silk98,hopkins06,booth09,cattaneo09}.
Therefore, the early SFR is overestimated, 
while the suppression of SFR in small galaxies is underestimated,
resulting in excessive early star formation prior to $z \sim 3$, 
by a factor of order 2.
As a result, the typical gas fraction and SFR at $z \sim 2$ are 
lower by a factor of $\sim 2$ than the average observed values in 
star-forming galaxies \citep{cdb10,daddi10,tacconi10}.

\smallskip
Furthermore, the simulated galactic mass outflow rate is only a fraction of  
the SFR, 
\adb{where the mass loading factor ranges from zero to unity with
an average $\eta \sim 0.3$ at $0.5\Rv$},
not reproducing some of the observed strong
outflows with mass loading factors of order unity and above 
\citep{steidel10,genzel11,dk13}.
This leads to a stellar fraction of $\sim 0.1$ within the virial radius,
a factor of $\sim 2-3$ higher than the observationally indicated value
\citep[e.g.][]{perezgonzalez08,behroozi13}.
These inaccuracies in the SFR, feedback and outflows introduce 
a limitation on the generality of our testing.
Nevertheless, this situation has the advantage that one can 
test the toy models for cosmological accretion and the response of the
galaxy to it without the extra complication of very strong feedback effects.
More accurate recipes for star formation and feedback are being incorporated
into simulations that we and others are now running, and they will be improved
further in the future. This will enable the next generation comparison to 
toy models.

\smallskip
\adb{
One should note that the sample of simulated galaxies is not a fair sample of 
halo or galaxy mass.
The halo masses are limited by selection to a rather narrow mass range at the 
target redshift $z=1$, and therefore to limited mass ranges at earlier 
redshifts, and the mass distribution at any redshift does not follow the 
$\Lambda$CDM halo mass function.
We do not attempt to follow the properties of a population of galaxies
as it evolves, partly because of the mass variation at $z=1$,
and partly because of the degrading of our sample at late times.
Instead, we treat each outputed snapshot independently of its history,  
assuming that it represents an (almost) arbitrary galaxy of an 
instantaneous mass $M$ at $z$.
Since the galaxies are of different masses at any given redshift,
we evaluate ``average" properties for the simulated galaxies focusing
on quantities that are only weakly dependent on mass within the spanned mass 
range, such as the specific accretion rate. 
For quantities that do have a significant mass dependence,
such as the accretion rate and the cumulative mass growth, we attempt to
scale out the mass dependence using the toy model itself prior to stacking 
the galaxies of different masses.
}

\smallskip
\adb{
It should be emphasized that we only follow the growth of the central galaxies
in the main-progenitor haloes of the final haloes selected at $z=1$.
The other progenitors, satellite and merging haloes (or galaxies),
are all considered here as part of the mass inflow onto the main progenitor.
The growth of the non-main-progenitor galaxies in the zoom-in simulations
is not studied here, partly because they represent only
10-20\% of the population of galaxies of the same mass, which mostly consists
of central galaxies, not satellites of more massive galaxies.
We do not attempt here an extension of the zoom-in simulations to
smaller central galaxies because the accuracy at the given resolution becomes
limited, and because the proper treatment of feedback effects in small
galaxies, where they dominate the evolution, is more demanding.
}

\section{Cosmological Accretion Rate}
\label{sec:accretion}

\subsection{Halos in the Einstein-deSitter Regime} 

We consider the $\Lambda$CDM cosmology 
in the Einstein-deSitter (EdS) regime. This is a useful approximation
at $z>1$, and it becomes more and more accurate at higher redshifts.
\adb{In our toy modeling, in the EdS regime, we adopt the following simple
relation between the expansion factor $a$ and the Hubble time $t$,}
\be
a=(1+z)^{-1} \simeq \left(\frac{t}{t_1}\right)^{2/3} \, ,
\label{eq:a_t}
\ee
\be
t_1 = \frac{2}{3} \omm^{-1/2} H_0^{-1} \simeq 17.5\Gyr \, .
\label{eq:t_1}
\ee
A comparison to the accurate expressions in the appendix, \se{useful},
reveals that 
this approximation overestimates $t$ at $z=1$ by only 5\%, and by 1.6\% at
$z=2$ (but by 28\% at $z=0$).  
The mean mass density in the Universe is
\be
\rho_{\rm u} = \rho_0 a^{-3} \, , \quad
\rho_0 \simeq 2.5\times 10^{-30} \,{\rm g\,cm}^{-3} \, .
\label{eq:rho}
\ee

\smallskip
Inspired by the spherical-collapse model and the virial theorem,
a halo is defined as a sphere about a density peak
that in the EdS regime 
encompasses a mean overdensity of $\Delta \simeq 200$ above
the cosmological background (see \equ{Delta} for a more accurate expression,
where $\Delta \simeq 207, 187$ and $178$ at $z=1,2$ and $z\gg 1$). 
The relations between virial mass $\Mv$, radius $\Rv$ and velocity $\Vv$ 
are thus
\be
\Vv^2 = \frac{G\Mv}{\Rv} ,
\quad \frac{\Mv}{(4\pi/3) \Rv^3} = \Delta \rho_u \, ,
\label{eq:vir0}
\ee
which lead to the approximate virial relations
\be
V_{200} \simeq M_{12}^{1/3} (1+z)_3^{1/2} ,
\quad
R_{100} \simeq M_{12}^{1/3} (1+z)_3^{-1} \, ,
\label{eq:vir}
\ee
where $M_{12}\equiv \Mv/10^{12}\msun$, $V_{200}\equiv \Vv/200\kms$,
      $R_{100} \equiv \Rv/100\kpc$, and $(1+z)_3\equiv (1+z)/3$.
As can be verified based on \equ{virial} in \se{useful},
\equ{vir} is accurate to a few percent at $z>1$. 

\smallskip
Based on \equ{vir} and \equ{a_t}, the halo crossing time $\tv$ scales with
the cosmological time $t$,
\be
\tv = \frac{\Rv}{\Vv} \simeq 0.14\, t \, .
\label{eq:tv}
\ee

\subsection{Toy Model: Accretion} 

We use the general term ``accretion" to refer to the
total inflow
\adb{into a distinct halo or its central galaxy}, 
including all the mass in dark matter, gas and stars 
and not distinguishing between smooth and clumpy components, i.e., including
all mergers.

\smallskip 
As reported in the Appendix, especially \se{acc_app},
the average specific accretion rate of mass into haloes 
of mass $M$ at $z$ can be approximated by an expression of the form
\be
\frac{\Mdot}{M} \simeq \ga\, M_{12}^\beta\, (1+z)^{\mu} \, . 
\label{eq:acc}
\ee
In the EdS regime $\mu \rar 5/2$. 
With $\beta = 0.14$ and the appropriate value of the normalization factor
$\ga$,
this approximation for the average of $M(z)$ was found to be 
\adb{a good match to cosmological N-body simulations}
to better than 5\% for $z>1$ 
\adb{\citep{neistein08b}},
while it becomes an underestimate of $\sim 20\%$ at $z=0$.
\adb{
Similar fitting formulae with slight variations in the values of the
parameters, valid in different ranges of mass and redshift,
were proposed by others
\citep{fakhouri08,genel08,genel10}.
}

\smallskip 
The power of $\mu=5/2$ can be simply understood from the following  
scaling argument based on the Press-Schechter (PS) and Extended-PS (EPS)
approximations of gravitational structure formation in cosmology
\citep{press74,bond91}.
A key element in the PS formalism is a self-invariant time variable, 
$\omega \propto D(a)^{-1}$, 
where $D(a)$ is the growth rate of linear density perturbations (\equ{da}). 
The self-invariance means that the growth rate of halo mass with respect 
to $\omega$ is independent of $\omega$, namely $dM/d\omega=const.$, 
which implies $\Mdot \propto \dot\omega$.
In the EdS regime, where $D(a) \propto a$ and $a \propto t^{2/3}$, this 
gives  $\Mdot \prop a^{-5/2}$. We therefore use $\mu=5/2$
for our toy model at $z>1$.
At lower redshifts, a better fit can be obtained with 
$\mu \simeq 2.4 \rar 2.2$ \citep{neistein06,neistein08b}.

\smallskip 
The small power $\beta$ reflects the log-slope of the fluctuation power
spectrum. The value $\beta \simeq 0.14$ fits well the Millennium simulation
merger trees for $\Mv$ in the range $10^{11}-10^{14}\msun$ \citep{neistein08b}.
According to EPS, it should be $\beta=(n+3)/6$, where $n \sim -2$ is the  
power index, $P(k) \propto k^n$, on the corresponding scales. This implies
that $\beta$ should be even smaller for smaller haloes.
\adb{Since $\beta$ is rather small, we approximate $\beta=0$ in the toy
model addressed in this paper.} 

\smallskip
The normalization factor $\ga$ is the specific accretion rate into a halo
of $\Mv=10^{12}\msun$ at $z=0$, or the inverse of the corresponding
accretion timescale, $\ga =\tau_{\rm in,0}^{-1}$ (\se{SS_disc_toy}).
As described in \se{useful},
the value of $\ga$ for haloes of $\Mv=10^{12}\msun$ 
has been estimated earlier to be $\ga \simeq 0.030 \Gyr^{-1}$,
by a fit to the systematic mass growth $M(a)$ in the Millennium cosmological 
simulation, 
scaled to the $\Lambda$CDM cosmological parameters assumed here 
\citep[following][]{neistein08b}.  
In particular, the quoted value of $\ga$ 
is for $\sigma_8 = 0.82$, and it scales as $\ga \propto \sigma_8^{-1}$.

\smallskip 
According to \equ{acc}, with $\ga \simeq 0.030 \Gyr^{-1}$, 
a halo of $10^{12}\msun$ at $z=2$ accretes baryons at a rate
$\dot{M}_{b,ac} \simeq 80 {\fb}_{0.17} \sy$, where ${\fb}_{0.17}$
is the universal baryon fraction in units of $0.17$.

\begin{figure}
\vskip 7.2cm
\includegraphics{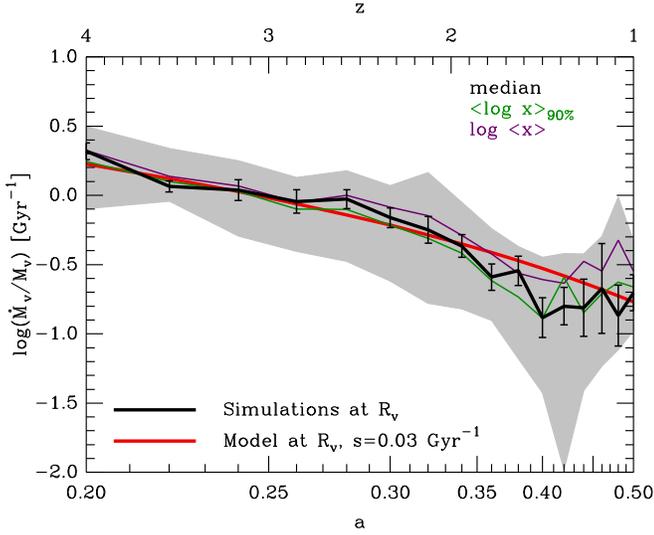}
\caption{
\adb{
Cosmological accretion of total mass:
specific mass inflow rate $\Mdot/M$ through the virial radius
as a function of time (expansion factor $a$ and redshift $z$).
Shown at each time step 
are the median (thick black), the average (magenta),
and the average of the log within the 90\% percentiles (green)
over the sample of simulated galaxies.
The error bars estimate the error of the mean, and 
the shaded area marks the 68\% percentiles.
Shown in comparison is the toy-model prediction,
\equ{acc_toy} with $\ga = 0.030 \Gyr^{-1}$ (thick smooth red).
The toy model provides a reasonable fit over the whole redshift range $z=4-1$
for the median and for the averages.
}
}
\label{fig:total_sMdot_rvir}
\end{figure}

\begin{figure}
\vskip 7.2cm
\includegraphics{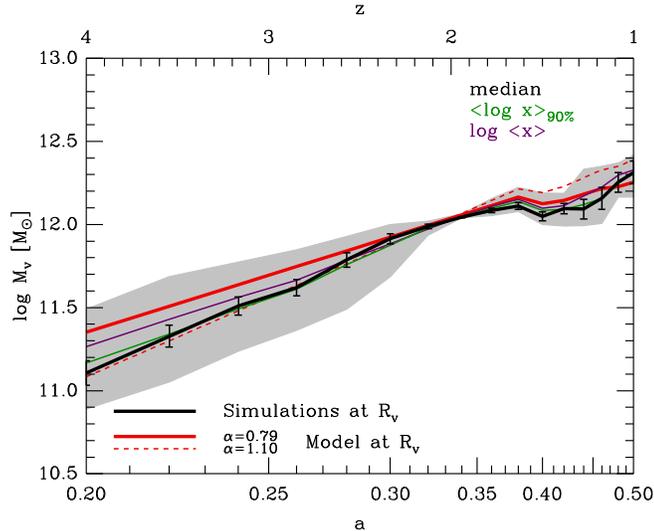}
\caption{
Cosmological accretion of total mass:
growth of virial mass.
Shown are the median (thick black), the average (magenta),
and the average of the log within the 90\% percentiles (green)
over the simulated galaxies.
The mass of each galaxy $M(z)$ has been scaled before stacking
by $\med[M(z=2)]/ M(z=2)$ (see text).
Shown in comparison is the toy model prediction,
\equ{mass_toy} with $\alpha=0.79$, 
normalized like the simulations at $z=2$ (thick smooth red).
The toy model is a reasonable approximation over the whole redshift range.
\adb{
Also shown is the toy model with $\alpha = 1.1$ (dashed red), which fits 
the median in the range $z = 4-2$.   
}
}
\label{fig:total_M_rvir}
\end{figure}

\begin{figure}
\vskip 7.2cm
\includegraphics{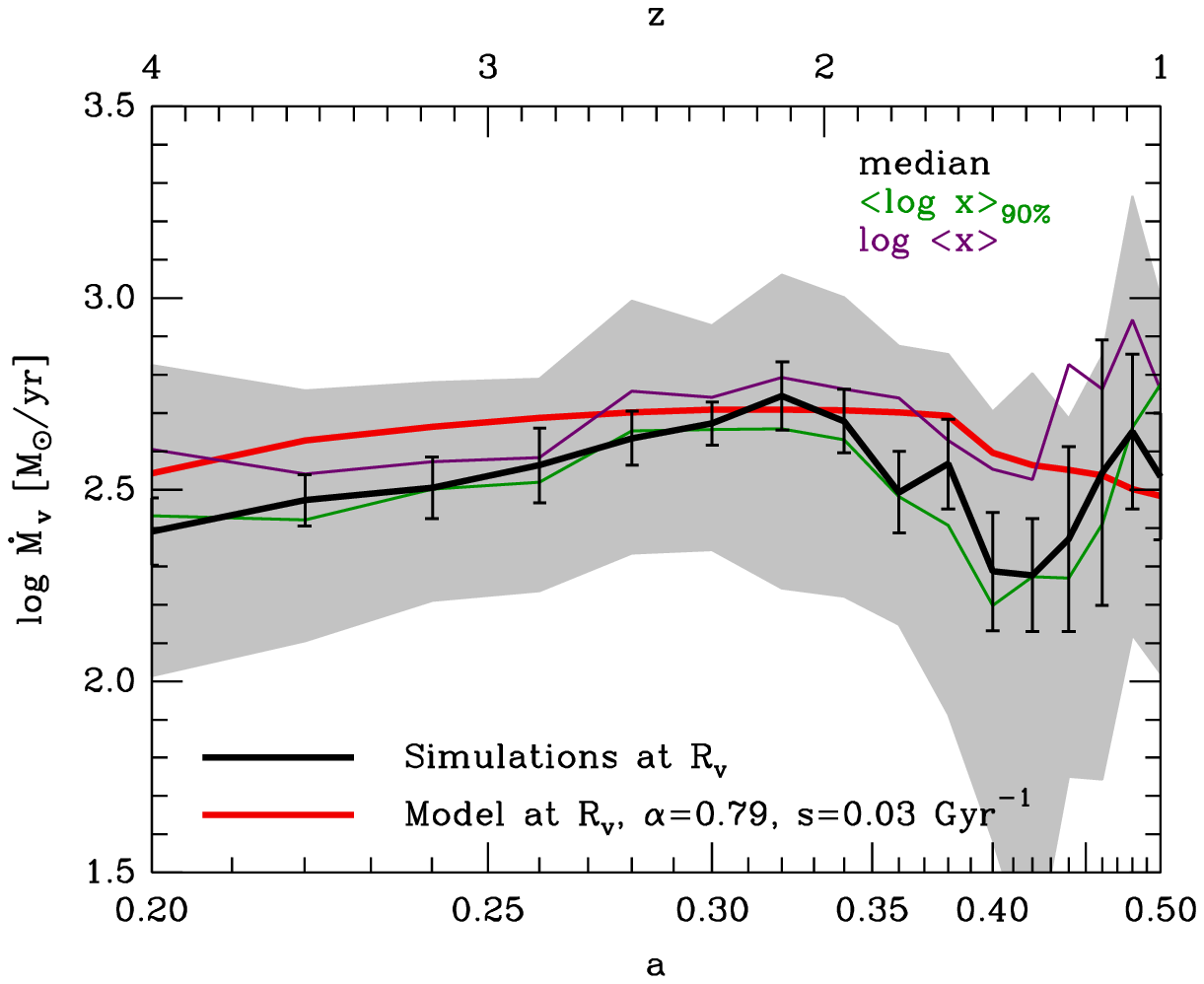}
\caption{
Cosmological accretion of total mass:
mass inflow rate $\Mdot$.
Shown are the median (thick black), the average (magenta),
and the average of the log within the 90\% percentiles (green)
over the simulated galaxies.
$\Mdot(z)$ of each galaxy has been scaled before stacking
by $\med[M(z=2)]/ M(z=2)$ (see text).
Shown in comparison is the toy model prediction,
\equ{Mdot_toy} with $\alpha=0.79$ and $\ga=0.030 \Gyr^{-1}$,
normalized like the simulations at $z=2$ (dashed red).
The model reproduces the stacked simulation results to $0.1-0.2$ dex.
}
\label{fig:total_Mdot_rvir}
\end{figure}

\smallskip
In a simple toy model 
\adb{that we address here}, 
valid for massive galaxies at $z>1$,
we ignore the weak mass dependence in the average specific
accretion rate, $\beta \rar 0$, and simplify \equ{acc} to
\be
\frac{\Mdot}{M} \simeq \ga\, (1+z)^{5/2} \, ,
\quad \ga \simeq 0.030\,\Gyr^{-1}\, .
\label{eq:acc_toy}
\ee
This can be simply integrated to a growth of halo mass as a function of $z$,
where the mass at some fiducial redshift $z_0$ is given to be $M_0$,
\be
\Mv = M_0\, e^{-\alpha\, (z-z_0)} \, ,
\quad \alpha = (3/2)\,\ga\, t_1 \simeq 0.79 \, .
\label{eq:mass_toy}
\ee
\adb{
This functional form was indeed found to be a good fit to the halo growth 
in earlier cosmological $N$-body
simulations of lower resolution \citep{wechsler02,neistein08b}.
}

\smallskip 
Note from \equ{acc_toy} and \equ{mass_toy}
that the accretion rate into a given halo as it grows is
\be
\Mdot(z) \simeq \ga\, M_0\, e^{-\alpha\, (z-z_0)}\, (1+z)^{5/2} \, .
\label{eq:Mdot_toy}
\ee
This average accretion rate into a given halo as it evolves
does not vary much in time over an extended cosmological period.
For $\alpha=0.79$, this rate has a maximum at
$z = 2.5/\alpha-1 \simeq 2.2$,
and it varies by less than a factor of 2 in the range $z\sim 0.3-5$.
The maximum average baryon accretion rate in the history of a halo of mass
$M_0=2\times 10^{12}\msun$ today, similar to the Milky Way,
is $\Mdot_{\rm max} \simeq 33 {\fb}_{0.17} \sy$.

\smallskip
While the expressions derived above are for the total accretion rate dominated
by dark matter, we suspect that the same expressions for the specific 
accretion rates could be valid for the specific accretion rate of baryons 
into the virial radius. This should be true when the baryons follow the 
total-mass inflow with a constant baryonic fraction, and as long as we 
tentatively ignore the baryonic mass loss from the haloes. 
Note that strong outflows can 
\adb{in principle} 
make the net baryonic accretion rate smaller than
the total accretion rate \citep[e.g.][]{faucher11b,vandervoort11b},
but the specific accretion rate may remain the same. 


\subsection{Simulations: Accretion}
\label{sec:sim_acc}

\subsubsection{Total accretion}
\label{sec:tot_acc}

In \figs{total_sMdot_rvir} to \ref{fig:total_Mdot_rvir},
we address the total mass inflow rate, dominated by the dark matter
component and including the baryons, gas plus stars.
\adb{
The accretion rate $\Mdot$ through a spherical boundary of 
radius $R$ during a timestep $\Delta t$ is taken to be the difference 
between the masses encompassed by the sphere of radius $R$ at the two snapshots 
defining the beginning and the end of the timestep, divided by $\Delta t$.
In most cases $R$ is either $\Rv$ or $0.1\Rv$, and in each of the snapshots
we use the actual value of $\Rv$ at that time. We verify in 
\se{penet_sim}, \fig{penet_10kpc}, that using a fixed radius does not make a
significant difference at $z>1$.}
\adb{
An alternative way to compute the instantaneous $\Mdot$ 
in a given snapshot would have been to sum over the cells or particles of 
mass $m_i$ and radial velocity $v_{r,i}$ within a shell of thickness $D$ 
about $R$, namely $\Mdot = \sum_i m_i\,v_{r,i}/D$.
As a sanity check, we have verified that when averaged over a timestep
this gives similar results to the former method.
}

\smallskip 
\adb{
We highlight in each figure the median over the galaxies within timesteps of 
$\Delta a = 0.02$, and also show the linear average, as well as the average 
of the log within the 90 percentiles. Because of the lognormal nature of the
distribution, the average of the log tends to be similar to the median, while
the linear average tends to be larger.
These stacked quantities are compared to the corresponding toy-model 
prediction.
}

\smallskip
\Fig{total_sMdot_rvir} shows the specific total mass inflow rate
$\Mdot/M$ through the virial radius.
The simulation results are compared to the toy-model prediction,
\equ{acc_toy}, with $\ga=0.030 \Gyr^{-1}$.
There is no other free parameter in the toy model for $\Mdot/M$.
\adb{
This comparison directly tests the predicted asymptotic systematic
time dependence, $\mu=5/2$, with the approximation that the specific inflow 
rate is independent of mass within the relatively narrow mass range spanned 
by the current sample. 
This mass independence allows a straightforward stacking of the galaxies of 
different masses without any scaling.
This makes the specific accretion rate the most robust quantity for the
comparison of the model with the simulations, and for determining the best-fit
normalization, the parameter $\ga$. 
We note, however, that a success in predicting the systematic evolution of 
$\Mdot/M$ does not guarantee an accurate match to the evolution of 
$\Mdot$ and of $M$, which could be both off by the same multiplicative factor. 
Another mismatch may occur because the average of $\Mdot/M$ is not necessarily
the ratio of the averages of $\Mdot$ and $M$.
}

\begin{figure*}
\vskip 5.3cm
\includegraphics{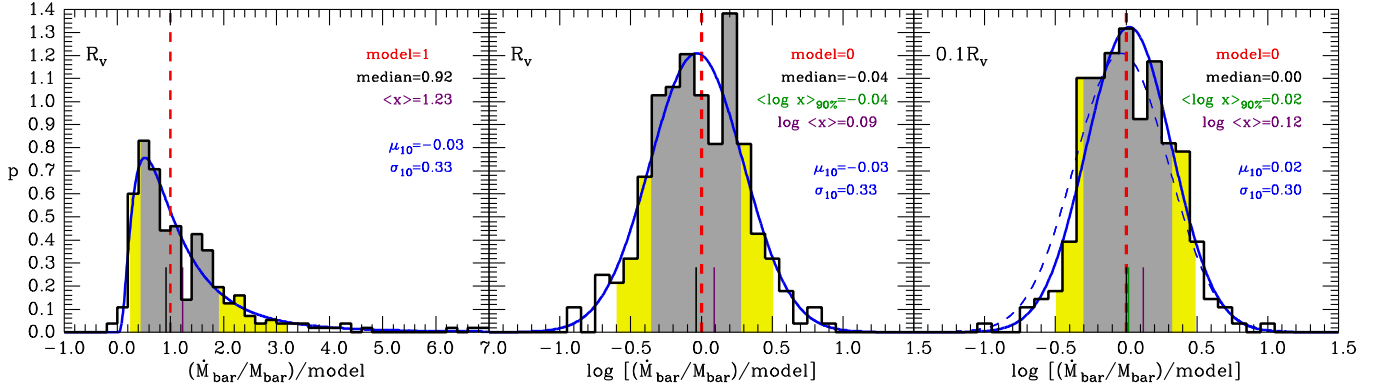}
\caption{
\adb{ 
Baryon accretion:
Distribution of specific accretion rate of baryons, $\Mdot/M$,
relative to the model prediction, as measured
at $\Rv$ (left linear, middle log) and at $0.1\Rv$ (right).
All snapshots of all galaxies are included, averaged within time steps of 
$\Delta a = 0.02$.
The distributions are close to lognormal, as can be seen from the lognormal
functional fits, with the associated log$_{10}$ mean ($\mu_{10}$) and 
standard deviation ($\sigma_{10}$) quoted.
The dashed Gaussian in the right panel is the fit at $\Rv$ from the middle
panel.
The scatter represents variations among the galaxies and along the history
of each individual galaxy (see \fig{baryon_acc_rvir_indiv}).
Marked by vertical bars are the median and linear average. 
The average of the log within the 90\% percentiles almost
coincides with the median.
The gray and yellow shaded areas denote the 68 and 90 percentiles 
about the median.
The negligible tail of negative values (excluded from the log plot)
corresponds to two rare cases.
The few percent tail of snapshots with high accretion rates
corresponds to mergers.
}
}
\label{fig:acc_Rv_histogram}
\end{figure*}

\smallskip
\adb{
One can see in \fig{total_sMdot_rvir} that, in general,
the model provides a good fit to the systematic redshift dependence in the
simulations, and that $\ga=0.030 \Gyr^{-1}$ gives a proper normalization.
The model matches the median (and log average) of the simulations well in
the redshift range
$z=4-2$, where our sample is complete and where we focus our analysis.
The model is a slight overestimate of the median in the range $z =1.8-1.3$,
with deviations at the $2\sigma$ level.
As expected, the linear average is somewhat higher than the median at every
time, and it is approximately matched by the model throughout the whole
redshift range $z=4-1$, though the average is slightly above the model at
$z=2.6-2.0$.
}

\smallskip
\Fig{total_M_rvir} shows the corresponding growth of total virial mass in time.
In order to stack all the galaxies, the mass of each galaxy at $z$
has been scaled by the inverse of the ratio of its mass at $z=2$
to the median mass at $z=2$,
namely $M(z)$ is multiplied by
$\med[M(z=2)]/M(z=2)$.
For the stacking at a given $z$,
the average $\la M(z=2)\ra$ is computed for the sample
of galaxies that is averaged over at $z$,
namely the whole sample of 27 galaxies at $z \ge 2$,
and the gradually diminishing subsample from $z=2$ to 1.
The simulation results are compared to the toy
model prediction, \equ{mass_toy}, with $\alpha=0.79$,
corresponding to $\ga=0.030 \Gyr^{-1}$.
The model is normalized at $z_0=2$ to match the simulations' median,
$M_0 = \med[M(z=2)]$, as defined above.
This way of stacking yields by construction no scatter and perfect fit of
the model at $z=2$; the comparison thus tests the success of the model
at redshifts away from $z=2$.
We learn that this model is a reasonable approximation for the total mass
growth over the redshift range studied.
\adb{
The model overestimates the median by $\sim 0.2$ dex at $z=4-3$,
but it is less than $0.1$ dex above the linear average.
The figure demonstrates that a higher value of $\alpha$ (and $s$)
can fit better the average or the median in the range $z = 4-2$, 
where the simulation results are more reliable.
}

\smallskip
\Fig{total_Mdot_rvir} shows the corresponding absolute total mass inflow rate
$\Mdot$ at $\Rv$.
Motivated by \equ{Mdot_toy} where $\Mdot \propto M$,
the value of $\Mdot$ for each galaxy at every
$z$ has also been scaled before stacking by $\med[M(z=2)]/M(z=2)$
(similar to the scaling applied to $M(z)$).
The simulation results are compared to the toy
model prediction, \equ{Mdot_toy}, with $s=0.03 \Gyr^{-1}$ and $\alpha=0.79$.
The toy model is again normalized at $z_0=2$ to $M_0 = \med[M(z=2)]$.
Indeed, the value of $\Mdot$ does not vary much in time during the evolution of
a halo within this redshift range, because the growth of halo mass compensates
for the cosmological decline of specific inflow rate at a given mass.
The gradual diminishing of the sample between $z=2$ and $z=1$
is reflected in the shape of the model curve in that region.
\adb{
The model reproduces the general trend with time seen in the stacked
simulations.
It overestimates the median by $\sim 0.1-0.2$ dex at $z=4-2.7$, and even by
more at $z \sim 1.4-1.3$, but it matches well the linear average over the whole
range (except at $z < 1.3$ where the sample is very small).
}

\smallskip
Our results for the total accretion at $\Rv$
confirm that the toy model in its simplest form provides
a very useful approximation for the characteristic values of the three
quantities, $\Mdot/M$, $M$ and $\Mdot$, over the whole redshift range.
We see that the simplest toy model enjoys different levels of success
for the different quantities at different redshifts.
Somewhat different values of $\ga$ and $\alpha$ may provide the best fit
\adb{
to the median and to the average,
}
to each of the three quantities, and at different redshift ranges.
There are several reasons for this.
\adb{
First, the lognormal nature of the distribution of $\Mdot$ makes the average
larger than the median.
}
Second, the three noisy quantities are not linearly related to each other,
and therefore their averages do not simply relate to each other.
Third, the toy model in \equ{acc_toy} and \equ{mass_toy}
slightly deviates from the more accurate approximation \equ{acc},
where $\beta \simeq 0.14$, which itself is only an approximation.

\begin{figure}
\vskip 7.2 cm
\includegraphics{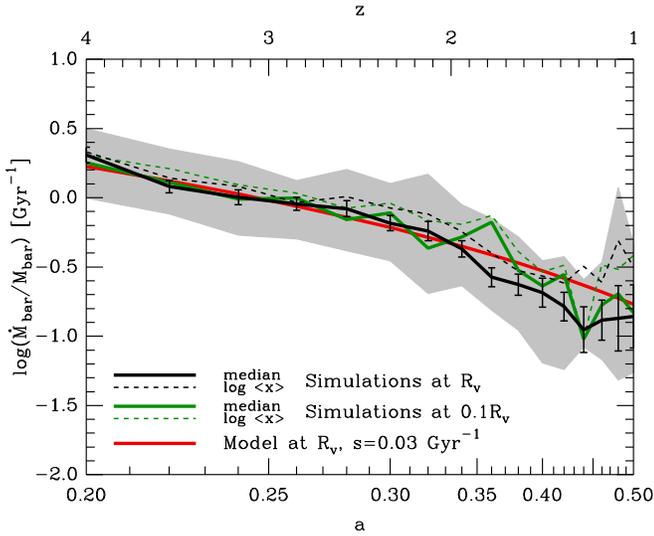}
\caption{
Baryon accretion:
specific inflow rate and penetration.
\adb{
Shown are the median (thick solid) and the average (thin dashed) of $\Mdot/M$ 
through $\Rv$ (black) and through $0.1\Rv$ (green).
At $\Rv$ it is analogous to the total rate in \fig{total_sMdot_rvir}.
The model prediction for $\Rv$, \equ{acc_toy} with $\ga=0.030 \Gyr^{-1}$ 
(smooth solid red), is indeed a good approximation at $\Rv$, 
but it also provides 
a good approximation for the specific accretion rate at $0.1\Rv$.
}
}
\label{fig:baryon_acc_rvir}
\end{figure}

\begin{figure}
\vskip 7.2cm
\includegraphics{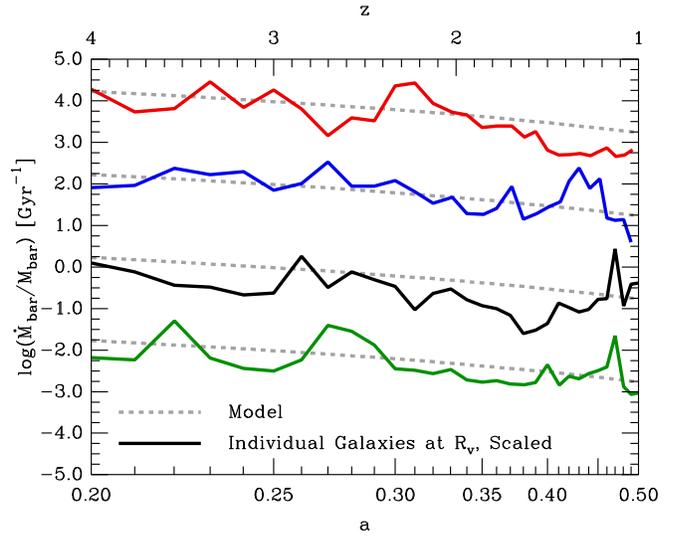}
\caption{      
Baryon accretion:
specific inflow rate at $\Rv$ for four individual galaxies
(that enter the average in \fig{baryon_acc_rvir}). 
The black curve second from bottom is normalized properly,
and the other curves are shifted by 2 dex relative to each other.
The simulations are compared to the toy model prediction,
\equ{acc_toy} with $\ga=0.030 \Gyr^{-1}$ (dotted).
This figure illustrates the scatter due to variations among the galaxies
and due to variations along the history of each individual galaxy.
}
\label{fig:baryon_acc_rvir_indiv}
\end{figure}

\subsubsection{Accretion of baryons}
\label{sec:bar_acc}

\begin{figure}
\vskip 7.2cm
\includegraphics{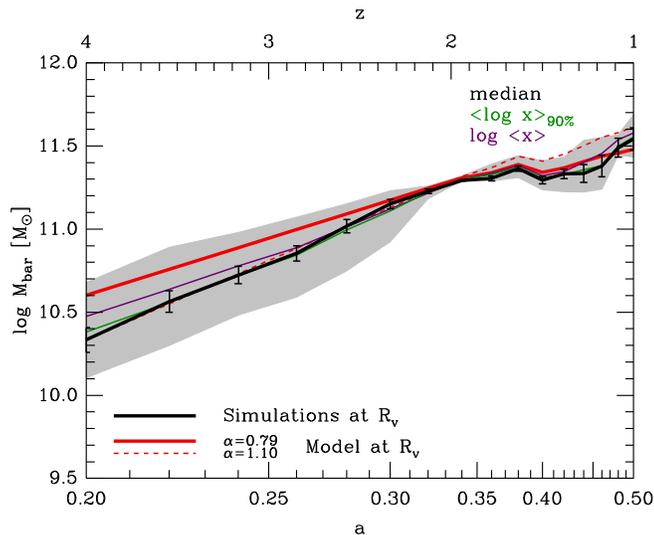}
\caption{
Baryon accretion:
growth of mass within $\Rv$, analogous to the total mass in
\fig{total_M_rvir}.
The toy model with the adopted value of $\alpha$ \equ{mass_toy}
is a reasonable approximation over the whole redshift range.
\adb{
Also shown is the toy model with $\alpha = 1.1$ (dashed red), which fits 
the median in the range $z = 4-2$.
}
}
\label{fig:baryon_growth}
\end{figure}

\adb{
The baryonic inflow through the virial radius in the simulations 
consists of stars and gas in comparable fractions,
with the stellar fraction increasing with time. Most of the stars are in
small merging galaxies, associated with mini-minor mergers of mass ratio 
smaller than 1:10 \citep{dekel09}. These mergers do not have a substantial
global dynamical effect on the disc --- they tend to join the disc as part
of the cold streams and grow the cold disc component that develops the violent
disc instability. On the other hand, the incoming gas component provides much
of the fuel for new star formation.
Recall that the stellar fraction at $z\sim 2$ is considered to be an
overestimate in the current 
simulations, by a factor of order two, because of the high efficiency assumed 
for star formation at high redshift \citep{cdb10}. 
We therefore focus on the more robust total baryonic accretion, and do not 
seek an accuracy of better than 
a factor of two when dealing with the gas accretion alone, e.g. in \se{SS} and
\se{inflow}.
}

\adb{\Figs{acc_Rv_histogram} to \ref{fig:baryon_growth}}
address the accretion of baryons through the virial radius.
The calculations for the baryonic accretion are analogous to those described in 
\se{tot_acc} for the accretion of total mass.
The first three figures refer to the specific rate, where
\adb{\fig{acc_Rv_histogram} shows the distribution over all snapshots,}
\fig{baryon_acc_rvir} shows the \adb{systematic} time evolution over the
simulations compared to the toy model prediction \equ{acc_toy},
and 
\fig{baryon_acc_rvir_indiv} shows the same for four individual galaxies.
Then \fig{baryon_growth} refers to the baryon mass growth in the simulations
compared to the model prediction \equ{mass_toy}.

\smallskip 
A comparison of \figs{baryon_acc_rvir} and \ref{fig:baryon_growth}
to \figs{total_sMdot_rvir} and \ref{fig:total_M_rvir} 
shows a strong similarity, implying that the baryons follow the dark matter 
in its streaming into the virial radii. 
This implies that the outflows 
through the virial radii are indeed rather small, 
at least on average in this suite of simulations.

\smallskip 
\adb{
In \fig{acc_Rv_histogram} we learn that the distribution of specific baryonic
accretion rates at $\Rv$ resembles a lognormal distribution.
This is very similar to the distribution of the total accretion rates,
which has been seen earlier to resemble a lognormal distribution
\citep{neistein08a}.
The median and averages of the ratio of $\Mdot/M$ to the model prediction
are close to unity, indicating that the toy model
provides a good approximation for the systematic behaviour.
The median and the very similar average of the log are at $0.92$,
and the linear average is $1.23$, somewhat larger as expected from a lognormal 
distribution. This is an argument for using the median or the average of the 
log to represent the systematic behaviour of the stacked sample.
The scatter of $\pm 0.33$ dex indicates that the specific accretion rate in 
different galaxies 
or at different times can deviate significantly from the median or averages. 
Small negative $\Mdot$ values, namely net outflows,
are measured in two rare cases, reflecting a combination of 
satellites moving out of the halo and gas outflows.
The extended tail at large accretion rates corresponds to mergers.
}

\smallskip 
\adb{
\Fig{baryon_acc_rvir} shows that, similar to the case of total accretion,
the toy model of \equ{acc_toy} provides
a good fit to the stacked simulation results at $\Rv$.
With $\ga = 0.03 \Gyr^{-1}$, it is an excellent match to the median at 
$z=4-2$. The model is a slight overestimate of the median at $z=1.8-1.0$,
at the level of $0.1-0.2$ dex, or 2-sigma in terms of the error 
of the mean in a few points.
With this normalization, the average is slightly higher than the model at 
$z=4-2$, and is matched well at $z = 1.8-1.2$.
}

\smallskip
\Fig{baryon_acc_rvir_indiv} shows the specific baryon accretion rate at $\Rv$
for four individual galaxies. It shows separately the scatter due to 
variations among the galaxies and the scatter due to variations along 
the history of each individual galaxy. Although some galaxies seem to deviate
systematically from the model throughout their histories (e.g. the second from
bottom case
tending to be lower than the model at most times), most galaxies fluctuate
about the model prediction. The typical scatter along the histories of  
individual galaxies is comparable to the variations from galaxy to galaxy.

\smallskip
\adb{
\Fig{baryon_growth} addresses the baryon mass growth within $\Rv$.
In analogy to \fig{total_M_rvir}, the galaxies are scaled before stacking
to the same value of the median at $z=2$.
The model overestimates the median by 0.2 dex at $z \sim 4$, but it
overestimates the average only by 0.1 dex at that redshift.
A higher value of $\alpha$ (and $s$) matches the growth curve better in the
range $z=4-2$.
}

\section{Penetration to the inner halo}
\label{sec:penetration}

\subsection{Toy model: Penetration}

\begin{figure}
\vskip 7.2cm
\includegraphics{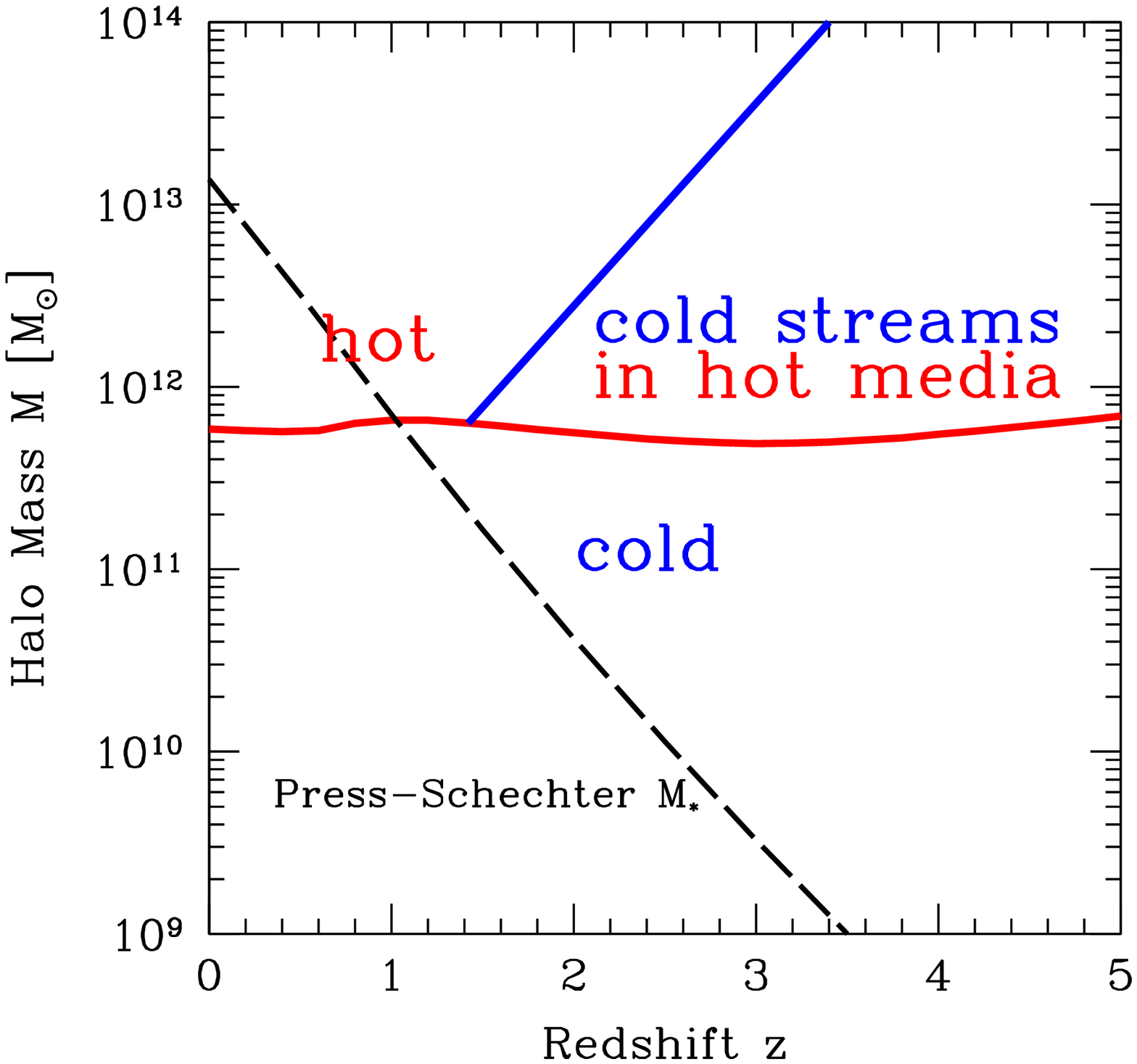}
\caption{
Predicted penetration of cold gas streams into the halo centre
as a function of halo mass and redshift. This schematic diagram 
\citep[reproduced from][]{db06,dekel09}, is based on analytic 
spheri-symmetric calculations \citep{db06,dekel09}.
While a virial shock is expected to be present at and above
$\Mv \sim 10^{12}\msun$, at high redshift the gas is expected to penetrate
through it in cold streams along the filaments of the cosmic web.
}
\label{fig:acc_toy}
\end{figure}

We expect, based on the reasons outlined below, that at $z >1$ 
a very large fraction of the baryons that enter the halo at the
virial radius efficiently penetrate into the galaxy at the halo centre.
Below a critical halo mass of $\sim 10^{12}\msun$, a stable virial shock 
encompassing an extended hot gas medium cannot be supported because the
radiative cooling rate is too high \citep{bd03,keres05,db06},
so the instreaming of cold gas is unperturbed.
A stable virial shock is likely to develop in more massive haloes, 
and it may shut down the cold accretion in sufficiently massive haloes
at low redshifts,
but at $z>1$, most of the accreted gas is expected to be in cold streams that 
penetrate through the hot halo medium deep into the inner halo 
\citep{ocvirk08,dekel09,keres09}, as illustrated in \fig{acc_toy}
\citep[based on][]{db06}.
At high redshift, in massive galaxies that represent high-sigma peaks
in the density fluctuation field,
the gas density in these streams is high compared to the 
mean gas density in the
halo, as they follow the narrow dark-matter filaments of the cosmic web. 
The high density enhances the radiative cooling rate in the streams
and prevents the development of pressure that could support a stable virial
shock in the streams. 
As a result, the streams penetrate through the halo with a mass inflow 
rate that is rather constant with radius \citep{dekel09}. 
They flow in at a constant inflow velocity that is comparable to the
virial velocity, and they keep cold at $\gsim 10^4$K while dissipating 
the gained gravitational energy into cooling Lyman-alpha radiation
\citep{loeb09,goerdt10,fumagalli11}.\footnote{Note that SPH simulations, 
which tend to underestimate the dissipation rate 
in the cold streams, may yield somewhat different results
\citep{faucher10,vandervoort11a,nelson13}.} 

\subsection{Simulations: Penetration}
\label{sec:penet_sim}

\begin{figure}
\vskip 7.2cm
\includegraphics{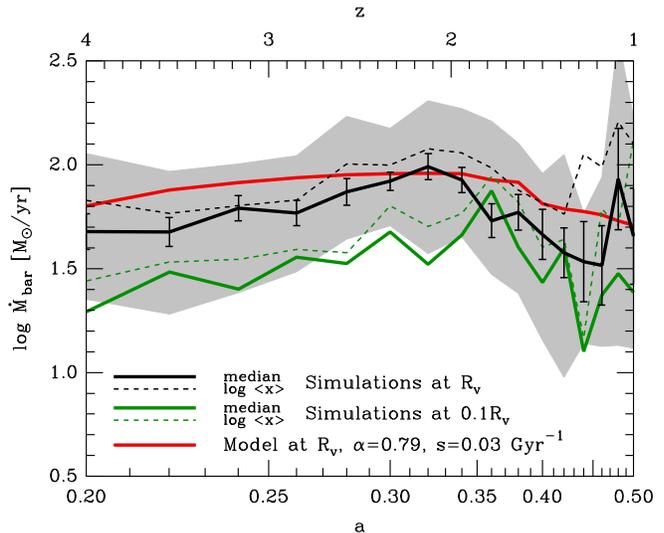}
\caption{
Baryon accretion: 
inflow rate and penetration.
Shown are the median (thick solid) and the average (thin dashed) of $\Mdot$
through $\Rv$ (black) and through $0.1\Rv$ (green).
At $\Rv$ it is analogous to the total rate in \fig{total_Mdot_rvir}.
The model prediction for $\Rv$, \equ{Mdot_toy} with $\alpha=0.79$, $\ga=0.030
\Gyr^{-1}$, and normalized like the simulations at $z=2$
(smooth solid red), is a good approximation at $\Rv$.
The penetration to the inner galaxy is $\sim 50\%$ at $z=4-2$, 
and higher at $z=2-1$.
}
\label{fig:penet_0.1Rv}
\end{figure}

\begin{figure}
\vskip 7.2cm
\includegraphics{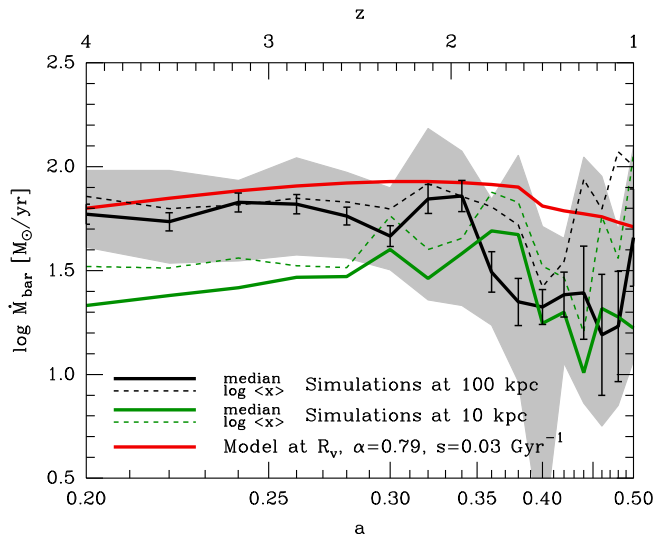}
\caption{
Baryon accretion:
inflow rate and penetration at fixed radii.
Same as \fig{penet_0.1Rv}, except that the evolving $\Rv$ and $0.1\Rv$ are 
replaced by fixed radii $100\kpc$ and $10\kpc$, respectively.
There is no qualitative change from \fig{penet_0.1Rv} at $z \geq 2$, 
indicating that the
artificial ``accretion" due to the growth of $\Rv$ is not a major part of 
the actual inflow rate into the central galaxy at high redshift. 
}
\label{fig:penet_10kpc}
\end{figure}

\Fig{baryon_acc_rvir}, which has been discussed above concerning the specific 
accretion rate of baryons at $\Rv$, also shows the 
specific accretion rate of baryons into the inner halo at $0.1\Rv$.
The stacked simulation results are compared to the model prediction for the
accretion rate at $\Rv$ at that time, \equ{acc_toy}.
\adb{
The right panel of \fig{acc_Rv_histogram}
}
shows the distribution of specific accretion rate
of baryons at $0.1\Rv$, over all the snapshots of all the galaxies,
normalized to the model prediction.
The distribution at $0.1\Rv$ should be compared to that shown in the left
panels for $\Rv$, 
and to the Gaussian fit that characterizes it.
We learn that the specific baryon accretion rate at $\Rv$ and at $0.1\Rv$
are rather similar, and well approximated by the toy model.
\adb{
In fact, the model is a good match to the median of the simulations at $0.1\Rv$
over the whole redshift range.
}

\smallskip 
The similarity between the specific accretion rate at $0.1\Rv$
and at $\Rv$ does not necessarily imply 100\% penetration through the halo,
as both $\Mdot$ and $M$ for the baryons could decrease between 
$\Rv$ and $0.1\Rv$ in a similar way. 
We conclude that \equ{acc_toy}, with $\ga = 0.030 \Gyr^{-1}$,
is also a useful approximation for the specific accretion rate of baryons 
into the galaxy itself.
This is for the distribution among galaxies of halo masses
that range from below $10^{11}\msun$ to above $10^{12}\msun$
in the redshift range $z=4-1$.

\smallskip  
\Fig{penet_0.1Rv} shows the absolute inflow rate $\Mdot$ for the baryons 
at $\Rv$ and at $0.1\Rv$,
and the corresponding toy model prediction at $\Rv$, \equ{Mdot_toy}.
The scaling to the average at $z=2$ before stacking the simulated galaxies 
and the corresponding normalization of the toy model is the same
as described in \se{tot_acc}. 
We first note in \fig{penet_0.1Rv} that the simplified toy model, 
\equ{Mdot_toy}, is a reasonable fit to the 
inflow rate of the simulations at $\Rv$ in the range $z=4-1$.
The model overestimates the median by 0.2 dex or less, and it fits the average
to within 0.1 dex (except at $z<1.3$ where the sample is very small).
Indeed, the value of $\Mdot$ does not vary much in time during the evolution of
a galaxy within this redshift range, because the mass growth compensates for 
the decline of specific inflow rate, as for \fig{total_Mdot_rvir}.
This fit is complementary to the good fits
provided by the model to the specific inflow rate and to the mass growth,
\figs{baryon_acc_rvir} and \ref{fig:baryon_growth}.

\smallskip  
\Fig{penet_0.1Rv} provides a new insight into the penetration of baryons
through the halo into the inner galaxy.
One reads that, in the range $z=4-2$, the baryonic inflow 
rate at $0.1\Rv$ is typically $\sim 50\%$ of the baryonic inflow rate at $\Rv$. 
In the range $z=2-1$ the penetration seems to be more efficient, though
the uncertainty is larger because the sample is smaller.
This massive penetration is in general agreement with earlier estimates 
based on the MareNostrum 
simulation with lower resolution \citep{dekel09,danovich12} and with the
high SFR observed in massive galaxies at $z \sim 2$ \citep{genzel08,genzel11}.
If the SFR follows the gas accretion rate into the disc (see \se{SS}), 
the SFR can in principle be a fixed significant fraction of the instantaneous 
overall baryon accretion rate into the virial radius. Of course, 
internal processes within the disc may suppress the SFR at early times
($z>3$, say) and less massive haloes ($\Mv < 10^{10}\msun$, say) 
\citep[e.g.][]{kd12}, and thus accumulate gas for higher SFR in the more 
massive galaxies at later times (e.g., $z \sim 2$). 

\smallskip  
\adb{
One should recall the caveat associated with the significant fraction of
stellar mass in the inflow through the halo into the central galaxy.
This may be partly responsible for the more efficient penetration seen at 
$z<1.8$, and it reminds us not to assume accuracy of better than a factor of
two when addressing the incoming gas and the resultant star formation rate.
}

\smallskip 
The radii of spheres through which the inflow rate has been considered 
so far were the virial radius for the outer halo, and a fixed fraction of it 
($0.1\Rv$) for the galaxy at its centre. This is in order to allow us to
consider self-similar radii that follow the halo and the galaxy as they grow.
However, the growth of $\Rv$ in time (which can be deduced from the virial
relations, \equ{vir}, and the toy model for halo mass growth, \equ{mass_toy})
is responsible for part of the mass growth within $\Rv$, which may not be
associated with actual inflow \citep{diemer12}. In order to quantify this
effect, \fig{penet_10kpc} shows the inflow rate through spheres of fixed
radii, $100\kpc$ and $10\kpc$, replacing $\Rv$ and $0.1\Rv$ of
\fig{penet_0.1Rv}. A comparison between the two figures indicates no
significant differences between the absolute values of $\Mdot$ at $z=4-2$,
and similar conclusions regarding the penetration efficiency. 
Similar results are obtained for the specific accretion rate.  
We conclude that, at least at $z \geq 2$, the vast majority of the inflow
measured through $\Rv$ or a fixed fraction of it is true inflow, 
and therefore proceed with the analysis using the radii that grow
self-similarly in time, $\Rv$ and $0.1\Rv$.

\section{Disc Size}
\label{sec:spin}

\subsection{Toy Model: Disc Size versus Virial Radius}

We assume that the characteristic disc radius scales with the halo
virial radius via a contraction factor $\tlambda$,
\be
\Rd \equiv \tlambda\, \Rv \, .
\label{eq:Rd}
\ee
We also assume that the characteristic circular velocity of the disc scales
with the virial velocity,
\be
\Vd \equiv \tnu\, \Vv \, .
\label{eq:nu}
\ee
Given that the halo is roughly an isothermal sphere, 
the value of $\tnu$ is expected to be of order unity, and we sometimes adopt 
$\tnu \simeq \sqrt{2}$, which is a good approximation for the Milky Way. 
The value of $\tlambda$ can be estimated in the common model 
where the galactic disc is assumed to form by dissipative gas contraction 
within the dark-matter halo \citep{fe80,mmw98,bullock01_j}.
If one assumes that the original specific angular momentum of the gas, $j$, is 
similar to that of the dark matter in the virialized halo, 
and if $j$ is conserved during the gas contraction, then 
\be
\tlambda \simeq \sqrt{2}\, \tnu^{-1} \lambda' \, ,
\ee
where $\lambda' \equiv j/(\sqrt{2}\Rv\Vv)$ is the halo spin parameter as 
defined by \citet{bullock01_j}.
Based on N-body simulations and tidal-torque theory, the halo spin parameter is
assumed to have a constant average value independent of mass
or time, $\lambda' \sim 0.035$ \citep[e.g.][]{bullock01_j},
implying that $\tlambda$ is constant and of a similar value.  

\smallskip
The disc dynamical time is then
\be 
\td = \frac{\Rd}{\Vd} \simeq \tnu^{-1} \tlambda\, \frac{\Rv}{\Vv}
\simeq 0.0071\, \tnu^{-1} \tlambda_{0.05}\, t \, ,
\label{eq:td}
\ee
where $\tlambda_{0.05}\equiv \tlambda/0.05$,
and where we have used the relation of the virial crossing time to the
cosmological time $t$, \equ{tv}.

\smallskip
Note that the distribution of spin parameter among haloes is lognormal
with a standard deviation of half a dex \citep{bullock01_j}.
This would translate to a large scatter about the average $\Rd$
of \equ{Rd}.

\smallskip
Using \equ{Rd} and \equ{vir}, one can derive a
useful expression for the baryonic surface density in the disc,
\begin{eqnarray}
\Sigma\!\!&\!\!=\!\!\!\!\!& \frac{\Md}{\pi \Rd^2} \nonumber \\
&\!\!\simeq\!\!\!\!&1.3\!\times\!10^9 \msun\kpc^{-2}\, m_{{\rm d},0.1}\,
\tlambda_{0.05}^{-2}\, M_{12}^{1/3} (1+z)_3^2 \ , 
\label{eq:Sigma}
\end{eqnarray} 
where $\Md$ is the cold mass in the disc,
and $m_{{\rm d},0.1} \equiv m_{\rm d}/0.1$ refers to the baryonic mass 
in the disc relative to $\Mv$.

\begin{figure}
\vskip 7.0cm
\includegraphics{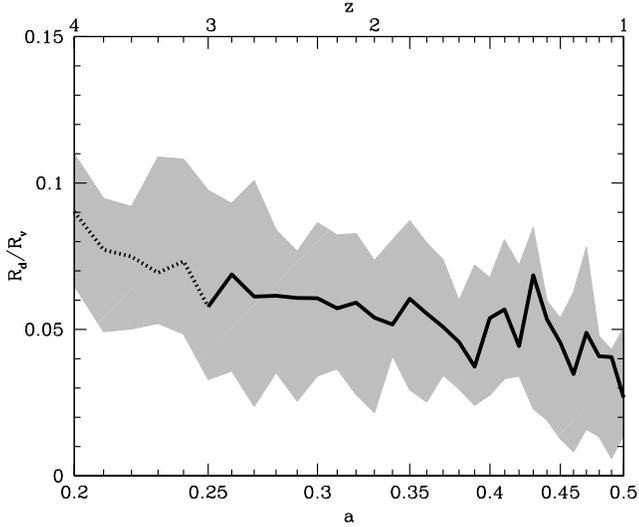}
\caption{
Ratio of disc radius to virial radius, $\tlambda$.
Shown is the median over the simulations (black curve) and the 68 percentiles
(shaded area) for the snapshots where there is net inflow, assumed to undergo
VDI.
The dotted line in the range $z=4-3$ reflects the large uncertainty in 
the disc radius in that regime. 
The ratio is in the ballpark of, and slightly higher than,
the expected value for the halo spin parameter, validating \equ{Rd}.
The average is slowly declining in time.
}
\label{fig:radii}
\end{figure}

\subsection{Simulations: Disc Size versus Virial Radius}

The disc radius $\Rd$ is determined in the simulations as described in
\citet{mandelker13}. 
In short, the disc is modeled as a cylinder whose $z$ axis coincides with the
spin axis of the cold gas within it ($T< 1.5\times 10^4$K, 
which is typically 97\% of the gas). 
We start with a large cylinder of radius $0.15\Rv$ and half-height $1\kpc$,
and iteratively converge on the cylinder of radius $\Rd$ that contains 
85\% of the gas mass within the large cylinder.
This determination of the disc radius is commonly in good qualitative agreement
with the radius one would have estimated by visual inspection of the face-on
surface density of the cold gas. Note, however, that 
this is only a crude approximation for the radius of the whole disc
including its stellar component, which could be dominant.

\smallskip
The disc radii as determined from the simulations at $z>3$ are rather  
uncertain and sometimes ill-defined. 
This is because the discs in these early times tend to be small and highly 
perturbed as the timescale for morphologically damaging mergers is comparable 
to and shorter than the disc dynamical time. 
We therefore prefer to de-emphasize the results that explicitly depend on the 
disc radius at $z>3$, e.g., in \fig{Mdot_inf} and \fig{Mdot_inf_acc} below.

\smallskip
\Fig{radii} shows the average and 68\% scatter
of the ratio $\Rd/\Rv$ over the simulated galaxies. 
In \equ{Rd}, a value of $\tlambda \simeq 0.05$ crudely fits the
1-sigma range of values from the simulations over the whole redshift range. 
However, there seems to be a systematic trend with
redshift, from a median of $\sim 0.06$ at $z=3-2$, through $0.05$ at $z=2-1.5$ 
down to $0.04$ at $z=1$. 
The indicated values of $\tlambda$ are somewhat larger than what is implied
from the spin parameter of dark matter haloes as estimated from 
cosmological N-body simulations, especially if $\tnu$ is larger than unity.
This is consistent
with observational indications at $z \sim 2$ \citep{genzel06}.
Indeed, the specific angular momentum of the baryons in the high-redshift
discs is likely to be higher than that of the dark-matter haloes 
(a) because of the way angular momentum is transported into the disc by 
cold streams from the cosmic web \citep{kimm11,pichon11,danovich12,stewart13}, 
and (b) because outflows tend to preferentially
carry away low-angular-momentum material \citep{maller02,brook11}.
We take the toy model for the disc radius and the associated estimate of
dynamical time, \equ{Rd} and \equ{td}, to be accurate to within a factor of
two.

\smallskip
Using abundance matching to determine halo virial radii, \citet{kravtsov13}
finds that the half-mass radii of today's galaxies of all morphological types, 
$R_{\rm half}$, are related to their today's halo radius $R_{200}$ as
$R_{\rm half} \simeq 0.015 R_{200}$. If the galaxy has formed roughly 
in its present size at $z \sim 2$, when the halo was a few times 
smaller in radius, the implied ratio then is consistent with our finding 
for $\tlambda$ in \fig{radii}.

\section{Mass Conservation: Steady-State of SFR and Gas Mass}
\label{sec:SS}

\subsection{Toy Model: Steady State}


Assuming mass conservation, the gas mass in the galaxy varies subject to
a source term and a sink term \citep[analogous to a ``bathtub",][]{bouche10},
e.g.,
\be
\dot{M}_{\rm g}=\dot{M}_{\rm g,ac}-\dot{M}_{\rm sf},
\label{eq:cont}
\ee
where $\dot{M}_{\rm g,ac}$ is the accretion rate of gas,
and $\dot{M}_{\rm sf}$ is the star formation rate.
We temporarily ignore potential gas outflow from the galaxy (to be incorporated
later).
We assume that the SFR density obeys a universal local volumetric law, 
$\dot{\rho}_{\rm sf} = \epsf \rho_{\rm g} /\tff$
with $\epsf \sim 0.02$ a constant efficiency and 
$\rho_{\rm g}$ the density of molecular gas in the star-forming region
\citep{kdm12}. 
Then, under certain circumstances, the overall SFR in the galaxy 
can be crudely assumed to be proportional to the total gas mass,
\be
\dot{M}_{\rm sf}=\frac{M_{\rm g}}{\tau_{\rm sf}} \, ,
\label{eq:sfr}
\ee
where $\tau_{\rm sf} = \tff/\epsf$. 

\smallskip
If $\dot{M}_{\rm g,ac}$ and $\tau_{\rm sf}^{-1}$ vary on a
timescale longer than $\tau_{\rm sf}$, the simple solution of \equ{cont} with
\equ{sfr} is
\be
\dot{M}_{\rm g} = \dot{M}_{\rm g,ac} e^{-t/\tau_{\rm sf}} \, , \quad
M_{\rm g} = \dot{M}_{\rm g,ac} \tau_{\rm sf}(1-e^{-t/\tau_{\rm sf}}) \, .
\label{eq:cont_solution}
\ee
After a transition period of order $\tau_{\rm sf}$, the solution relaxes to
a steady state solution 
\be
\dot{M}_{\rm g} \simeq 0 \, , \quad
M_{\rm g} \simeq  \dot{M}_{\rm g,ac} \tau_{\rm sf} \, , \quad
\dot{M}_{\rm sf} \simeq \dot{M}_{\rm g,ac} \, . 
\label{eq:ss}
\ee
The SFR sink term adjusts itself to match the
external source term $\dot{M}_{\rm g,ac}$. 

\smallskip
The value of $\tau_{\rm sf}$ can be related to the global disc
crossing time $\td$ by $\tff  = \fsf \td$, with $\fsf \sim 0.5$, 
assuming that star formation occurs in regions (partly bound clumps) 
where the overdensity is a few with respect to the mean density in the disc.
Using the estimate for $\td$ from \equ{td}, we obtain, for $\fsf \sim 0.5$
and $\epsf \sim 0.02$,
\be
\tau_{\rm sf} = \epsf^{-1}\, \fsf\,\td \sim 0.17\,t \, .
\label{eq:tausf}
\ee
At $z = 2$, where the Hubble time is $t \sim 3.25 \Gyr$, 
we have $\tau_{\rm sf} \sim 500-600 \Gyr$.

\smallskip
The validity of the solution \equ{cont_solution} to \equ{cont} 
depends on the timescales for
variation of $\dot{M}_{\rm g,ac}$ and of $\tau_{\rm sf}^{-1}$ compared to 
$\tau_{\rm sf}$.
Using \equ{acc_toy} and \equ{a_t} we obtain
\be
t_{\rm var}(\dot{M}_{\rm g,ac}) 
= \frac{\dot{M}_{\rm g,ac}}{\vert d \dot{M}_{\rm g,ac}/dt \vert}
\simeq \frac{t}{\vert (2/3)\,\alpha\, a^{-1} - 5/3 \vert} \, .
\ee
With $\alpha=0.79$ in \equ{mass_toy}, 
we get $t_{\rm var}(\dot{M}_{\rm g,ac})/t \simeq 1.0, 2.3, 12, 1.6$
for $z=4,3,2,1$ respectively. This means that 
$t_{\rm var}(\dot{M}_{\rm g,ac}) \geq t$ throughout the redshift range $z=4-1$,
so, based on \equ{tausf}, $t_{\rm var}(\dot{M}_{\rm g,ac}) \gg \tau_{\rm sf}$,
as required.
Similarly, using \equ{tausf}, the timescale for variation of 
$\tau_{\rm sf}^{-1}$ is  
\be
t_{\rm var}(\tau_{\rm sf}^{-1}) 
= \frac{\tau_{\rm sf}^{-1}}{\vert d \tau_{\rm sf}^{-1}/dt \vert}
= t \, ,
\ee
namely $\vert t_{\rm var}(\tau_{\rm sf}^{-1}) \vert \gg \tau_{\rm sf}$, as
required.
We thus expect the steady-state solution to be valid
through the whole period of interest here, and to be approximately valid
out to $z \sim 10$ and earlier.

\smallskip
A continuity equation of a similar nature, 
and its steady-state solution, were used with a variety of sink terms
in several recent studies
\citep{dsc09,bouche10,kd12,cdg12,gdc12,feldmann13}.
The sink term in \equ{cont} can be generalized to include other sink terms. 
For example, if there is an outflow at a rate
$\dot{M}_{\rm out} = f_{\rm out} \dot{M}_{\rm sf}$,
the second term of \equ{cont} is simply multiplied by the factor
$(1+f_{\rm out})$. Similarly, if \equ{cont} is applied only to the disc
component, instability-driven mass inflow within the disc adds
another sink term of a similar form, see \se{inflow}.
In these cases, $\tau_{\rm sf}$ in \equ{cont} is replaced by a smaller 
timescale $\tau$, which makes the steady-state solution an even better 
approximation.  

\begin{figure}
\vskip 7.2cm
\includegraphics{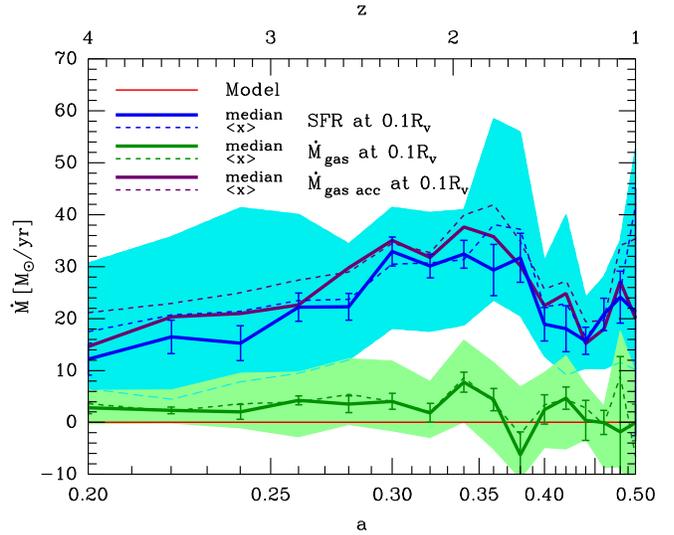}
\caption{
Overall steady state:
\adb{
Shown are SFR and $\Mdot_{\rm gas}$ within the sphere of radius $0.1\Rv$,
as well as the gas accretion rate $\Mdot_{\rm gas,acc}$ through that radius,
which is the sum of the former two.
Shown are the medians (solid thick) and the average (dashed thin)
over the simulated galaxies. 
The 68 percentiles (shaded area) and the errors of the mean (error bars)
are shown for $\Mdot_{\rm gas}$ and SFR.
The simulations evolve about the steady-state solution, \equ{ss},
where SFR$=\Mdot_{\rm gas,acc}$ and $\Mdot_{\rm gas} = 0$,
with a slow increase in the gas mass at the level of $\sim 10\%$ of the
SFR and the accretion rate.
}
}
\label{fig:sfr_acc}
\end{figure}


\begin{figure}
\vskip 7.2cm
\includegraphics{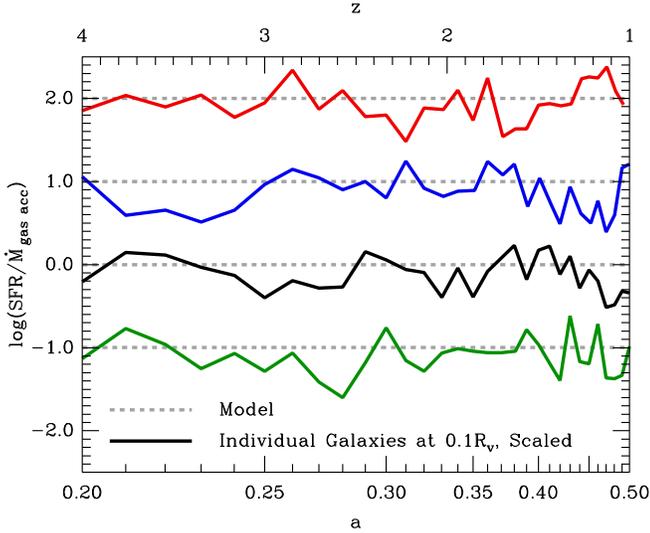}
\caption{
Overall steady state:
SFR versus gas accretion rate at $0.1\Rv$
for four individual galaxies.
The black curve second from bottom is normalized properly,
and the other curves are shifted by 2 dex relative to each other.
The simulations are compared to the toy-model steady-state solution,
\equ{ss}, where SFR$=\dot{M}_{\rm gas\,acc}$ (dotted).
This figure illustrates the scatter due to variations among the galaxies
and due to variations along the history of each individual galaxy.
}
\label{fig:sfr_indiv}
\end{figure}

\begin{figure}
\vskip 6.7cm
\includegraphics{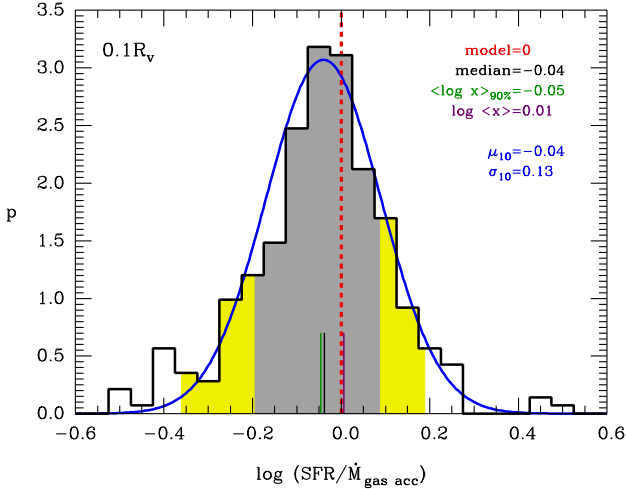}
\caption{
Overall steady state:
distribution of the ratio of SFR and gas accretion rate at $0.1\Rv$.
All snapshots of all galaxies are included, averaged within time steps of
$\Delta a = 0.02$.
The distribution is close to lognormal, as can be seen from the lognormal
functional fit, with the associated log$_{10}$ mean ($\mu_{10}$) and
standard deviation ($\sigma_{10}$) quoted.
The scatter represents variations among the galaxies and along the history
of each individual galaxy (see \fig{sfr_indiv}).
Marked by vertical bars are the median and linear average.
The average of the log within the 90\% percentiles almost 
coincides with the median.
The shaded areas denote the 68 and 90 percentiles about the median.
The SFR tends to be comparable to the gas accretion rate into the central
galaxy, with a median SFR$/\Mdot_{\rm gas\,acc} \simeq 0.91$ and an
average of $1.02$, compared to the model prediction of unity.
}
\label{fig:sfr_hist}
\end{figure}


\subsection{Simulations: Steady State}

\adb{
\Figs{sfr_acc} to \ref{fig:sfr_hist} verify the validity of the
steady state solution, \equ{ss}, in the simulations.
\Fig{sfr_acc} shows the SFR and the rate of change of gas mass, 
$\Mdot_{\rm gas}$, within the inner sphere of radius $0.1\Rv$.
As in \se{tot_acc}, 
these quantities were scaled before stacking to match the median mass at $z=2$.
The gas accretion rate through that radius, $\Mdot_{\rm gas,acc}$, 
is the sum of the two.
We see that the median and average of $\Mdot_{\rm gas}$ is indeed small, 
at the level of $\sim 10\%$ of the SFR or the accretion rate.
This corresponds to the SFR being close to the accretion rate, typically
only $\sim 10\%$ smaller.
}

\smallskip
\Fig{sfr_indiv} shows the ratio of SFR and $\Mdot_{\rm gas,acc}$ at $0.1\Rv$
for four individual galaxies, compared to the steady-state prediction of unity,
illustrating the variation along the history of each galaxy and among
the different galaxies. As in \fig{baryon_acc_rvir_indiv}, 
we see that the typical galaxy does not tend to show a systematic deviation 
of the SFR from the gas accretion rate into the disc along its history.
The scatter along the history of each individual galaxy is similar in the
different galaxies and comparable to the variations between different galaxies.

\smallskip
\adb{
\Fig{sfr_hist} shows the distribution of SFR$/\Mdot_{\rm gas,acc}$ at
$0.1\Rv$ over all the snapshots. It is close to lognormal, with a 
log$_{10}$ mean $\mu_{10} = -0.04$ and standard deviation $\sigma_{10} = 0.12$.
The linear average is only 2\% off unity.
We conclude that the SFR closely follows the gas accretion rate into the
$0.1\Rv$ sphere, as predicted by the steady-state solution, \equ{ss}.
While the scatter in this ratio is only at the level of $\sim 30\%$,
the much larger scatter in SFR seen in \fig{sfr_acc} indicates large
variations in both accretion rate and SFR between galaxies and/or along the
history of each galaxy.
}


\section{Inflow within the disc}
\label{sec:inflow}

\subsection{Toy Model: Disk Inflow}

Galactic discs at high redshift are expected to develop a 
gravitational disc instability with a Toomre instability
parameter $Q$ smaller than unity \citep{toomre64}.
The Toomre parameter can be expressed as \citep[e.g.][]{dsc09} 
\be
Q \simeq \frac{\sqrt{2}\Omega\sigma}{\pi G \Sigma} 
\simeq \frac{\sqrt{2}}{\delta} \frac{\sigma}{V} \, .
\label{eq:Q}
\ee 
The radial velocity dispersion $\sigma$ and the surface density $\Sigma$
refer to the cold component of the disc that participates in the disc 
instability (see the practical definition in the simulations in \se{inflow_sim}
below). 
We may refer to it as ``gas", but it also includes the cold 
young stars, and one should be careful to specify whether the latter are
included \citep[see a multicomponent analysis in][]{cdg12}. 
The angular velocity $\Omega$ and circular velocity $V$
refer to the characteristic disc radius $\Rd$, $V=\Omega \Rd$.
The constant $\sqrt{2}$ refers to a flat rotation curve 
(and it stands for $\sqrt{2(1+\theta)}$ where $V \prop r^\theta$).
The mass fraction $\delta$ is the ratio of mass in the cold component $\Md$ 
to the total mass within $\Rd$, $\Mt(\Rd)$,
the latter including gas, stars and dark matter,
\be
\delta = \frac{\Md}{\Mt(\Rd)} \, .
\ee

\smallskip 
The instability in high-redshift galaxies
is driven by the high cold surface density that reflects the 
high gas accretion rate and the high mean cosmological density at earlier 
times.  
We term this phase Violent Disk Instability (VDI), as the associated
dynamical processes occur on a galactic orbital timescales, as opposed to the
``secular" processes associated with disc instabilities at low redshifts.
The unstable disc tends to self-regulate itself in marginal instability
with $Q \simeq 1$, where $\sigma$ represents supersonic turbulence that 
provides the pressure while thermal pressure is negligible.
The turbulence tends to decay on a timescale comparable to the disc
dynamical time \citep{maclow99}, so it should be continuously powered
by an energy source that could stir up turbulence and maintain $\sigma$ 
at the level required for $Q \simeq 1$.
In the perturbed disc, which consists of extended transient features and 
massive compact clumps \citep{mandelker13}, 
gravitational torques drive angular momentum out
and cause mass inflow towards the centre, partly in terms of clump migration
\citep{noguchi99,bournaud07c,dsc09}, 
and partly in terms of off-clump inflow
\citep{gammie01,dsc09,bournaud11}.
This inflow down the potential gradient from the disc outskirts to its 
centre in turn provides the required energy for $Q \simeq 1$
\citep{krumholz10,cdg12,forbes12,forbes13}. 
The gas inflow rate ${\dot M}_{\rm g,in}$
can be estimated by equating this energy gain 
and the dissipative losses of the turbulence, 
\be
{\dot M}_{\rm g,in} V^2 \simeq \frac{\Mg \sigma^2}{\gamma \td} \, .
\label{eq:energy}
\ee
Here $\gamma  = (2/3)\gamma_{\rm g}^{-1}\gamma_{\rm \Phi}\gamma_{\rm dis}$,
where $\gamma_{\rm g}$ is the fraction of gas in the inflowing mass,
$\gamma_\Phi V^2$ is the energy gain per unit mass between the disc 
radius and the centre, and $\gamma_{\rm dis} \td$ is the turbulence 
decay timescale. The value of $\gamma$ is of order unity, and it 
could be as large as a few.
We thus obtain
\be
\frac{\dot{M}_{\rm g,in}}{\Mg} \simeq \frac{1}{\gamma\td} \frac{\sigma^2}{V^2} 
 \simeq \frac{1}{2\gamma\td} \delta^2 \, ,
\label{eq:Mdot_inf}
\ee
assuming $Q\sim 1$ in \equ{Q}.
Note that $\delta$ on the right-hand side refers to  the cold component, 
including young stars, while the left-hand side refers to the gas only.

\smallskip
Independent estimates based on the mechanics of driving the mass inflow by 
torques yield similar results for the inflow rate, to within a factor of two. 
Examples of such calculations are (a) an 
estimate of the rate of energy exchange by gravitational clump encounters 
\citep[][eqs.~7 and 21]{dsc09}, 
and
(b) an estimate of the angular-momentum exchange among 
the transient perturbations in a viscous disc 
\citep[][eq.~24]{gammie01,genzel08,dsc09}. 
In Appendix \se{df} 
we estimate that the disc evacuation rate by clump migration 
due to the dynamical friction exerted by the disc on the clumps 
is in the same ballpark as the 
above estimates of the inflow rate as long as $\delta \sim 0.2-0.3$
as in the steady-state solution for VDI discs \citep{dsc09}
and in our current simulations, \fig{delta}. 
One should emphasize that the inflow in the disc is a robust feature of
the instability, not limited to clump migration. 
Inflow at a rate comparable to \equ{Mdot_inf} is expected
even if bound clumps were to disrupt by stellar feedback in less than a 
migration time, as sometimes assumed  
\citep{murray10,genel12,hopkins12c}, though this scenario is not realistic based
on theoretical and observational grounds \citep{kd10,dk13}.

\smallskip
We comment in parentheses that
the cosmological streams that penetrate through the halo and feed the disc
are also potential drivers of turbulence \citep{dsc09,khochfar09,gdc12}, 
but this
requires strong coupling of the streams with the higher-density disc,
which demands that the streams largely consist of dense clumps. 
Being driven by an external source, this mechanism is not naturally 
self-regulating.

\begin{figure}
\vskip 7.0cm
\includegraphics{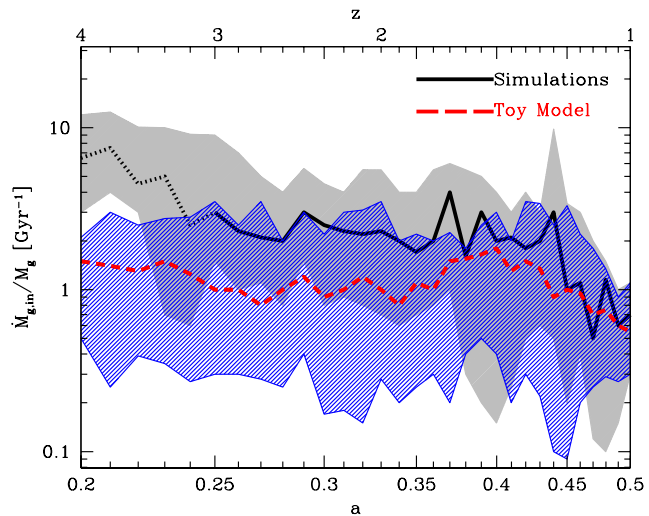}
\caption{
Inflow within the disc:
ratio of gas inflow rate and gas mass in the disc, testing \equ{Mdot_inf}.
Representing the left-hand side is the median over the simulations
(black curve) and the 68 percentiles (gray shaded area), limited to the
snapshots where there is net inflow, assumed to undergo VDI.
This is compared to the right-hand side of the toy model, \equ{Mdot_inf},
where $\delta$ and $\td$ are deduced from the simulations, with 
$\tnu =\sqrt{2}$
and $\gamma = 1$, showing the median (dashed red curve) and the 68 percentiles
(blue shaded area).
The median of the right-hand side typically underestimates that of the
left-hand side by a factor of $\sim 2$, but the fit is acceptable within
the scatter and the expected uncertainties, which are especially large at
$z>3$.
}
\label{fig:Mdot_inf}
\end{figure}

\begin{figure}
\vskip 7.0cm
\includegraphics{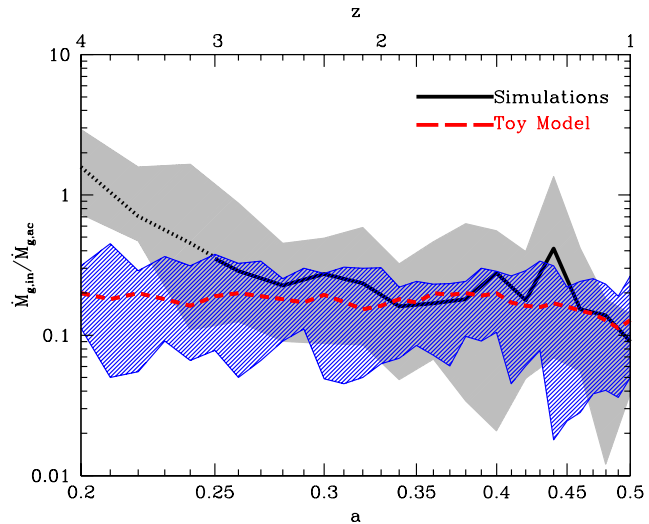}
\caption{
Inflow within the discs:
ratio of gas inflow rate in the disc and gas accretion rate into the disc
through a sphere of $0.1\Rv$, testing \equ{Mdot_inf_acc}.
Representing the left-hand side is the median over the simulations
(black curve) and the 68 percentiles (gray shaded area), limited to the
snapshots where there is net inflow, assumed to undergo VDI.
This is compared to the right-hand side of the toy model, \equ{Mdot_inf_acc},
where $\delta$ is computed from the simulations and with  
$\gamma = 1$, $\epsf =0.05$, and $\fsf=0.5$, 
showing the median (dashed red curve) and the 68 percentiles
(blue shaded area).
The fit is good except at $z>3$, where the uncertainty is large.
}
\label{fig:Mdot_inf_acc}
\end{figure}

\smallskip
Stellar feedback may or may not be a major direct driver of the
turbulence. There is potentially enough energy and momentum in supernovae,
radiative feedback and winds \citep{dsc09,bournaud10,kt12,kt13,dk13},
but it is not clear whether they are properly deposited within the disc.
Evidence against stellar feedback as a direct source of turbulence is in the
observation that the velocity dispersion is not tightly
correlated with proximity to star-forming clumps \citep{forster09,genzel11}. 
However, outflows can help boosting up $\sigma$ by simply lowering the
surface density of that gas that is supposed to be stirred up by a given energy
source \citep{gdc12}. 
Recall that the stellar feedback in the simulations used here is
underestimated.

\smallskip
According to \equ{ss}, the steady-state solution implies
$\Mg \simeq \dot{M}_{\rm g,ac} \tau_{\rm sf}$.
Combined with \equ{Mdot_inf} and $\tau_{\rm sf} = \fsf \epsf^{-1} \td$ from
\equ{tausf},
this gives in steady state
\be  
\frac{\dot{M}_{\rm g,in}}{\dot{M}_{\rm g,ac}}
\simeq \frac{\fsf}{\epsf\gamma} \frac{\sigma^2}{V^2}
\simeq \frac{\fsf}{2\epsf\gamma} \delta^2 \, .
\label{eq:Mdot_inf_acc}
\ee
Note that this expression is independent of $\td$.
The right hand side is of order unity, so the equation implies that the rate
of gas draining from the disc to the central bulge,
$\dot{M}_{\rm g,in}$,
is expected to be comparable to the rate of replenishment by
freshly accreting gas, $\dot{M}_{\rm g,ac}$. 
Since the steady-state solution, \equ{ss}, implied that
the SFR is also comparable to the gas accretion rate, we conclude that the
inflow rate in the disc is expected to be comparable to the SFR.  

\smallskip
The value of $\dot{M}_{\rm g,in} \Delta t$ can serve as a lower limit for
the growth of mass in the bulge during the period $\Delta t$.
The ratio of rates in \equ{Mdot_inf_acc} can be interpreted as an approximation
for the bulge-to-total baryonic mass ratio in the galaxy.

\smallskip
The gas mass fraction within the disc radius, $\delta_{\rm g}$, can be 
crudely estimated from the toy models as follows.
The gas mass is given by the steady-state solution, \equ{ss},
$\Mg \simeq \dot{M}_{\rm g,ac} \tau_{\rm sf}$. Expressing 
$\tau_{\rm sf} \simeq \fsf \epsf^{-1} \td$
and $\td \simeq 0.007 t$ from \equ{td}, we obtained 
$\tau_{\rm sf}/t \simeq 0.17$. 
We can then write $\Mg \simeq (\tau_{\rm sf}/t) \fg \fb \dot{M}_{\rm ac} t$,
where $\fg$ is the gas fraction in the baryon accretion, $\fb \simeq 0.17$
is the universal baryon fraction, and $\dot{M}_{\rm ac}$ is the total accretion
rate, such that $\dot{M}_{\rm ac} t \simeq \Mv$.
The total mass within the disc radius can be crudely estimated as the sum of the
baryonic mass and the dark mass, 
$M_{\rm tot} \simeq (\fb + \tlambda) \Mv$, assuming $M(r) \propto r$
within the halo.
We thus obtain for the gas fraction
\be
\delta_{\rm g} \simeq \frac{(\tau_{\rm sf}/t) \fg \fb}{\fb+\tlambda}
\lsim 0.1 \, .
\label{eq:delta_0.1}
\ee
If only a fraction of the baryons in the halo are within the disc radius,
the value of $\delta_{\rm g}$ would be larger accordingly.
When including the cold stars in the disc, the value of $\delta$ could be 
doubled or higher by a factor of a few, so we expect $\delta \sim 0.2-0.3$.

\subsection{Simulations: Disk Inflow}
\label{sec:inflow_sim}

\Fig{Mdot_inf} and \fig{Mdot_inf_acc} test the validity of the toy model
estimate for the inflow within the disc, \equ{Mdot_inf} and \equ{Mdot_inf_acc}
respectively.
For each equation, we compare the left-hand side as measured in the simulations
and the right-hand side as deduced from the toy model but also using quantities
that are measured from the simulations.

\smallskip
We adopt as our fiducial value for the dissipation timescale $\gamma=1$,
and the simulations can actually be used to determine the best-fit effective
value of $\gamma$.

\begin{figure}
\vskip 7.0cm
\includegraphics{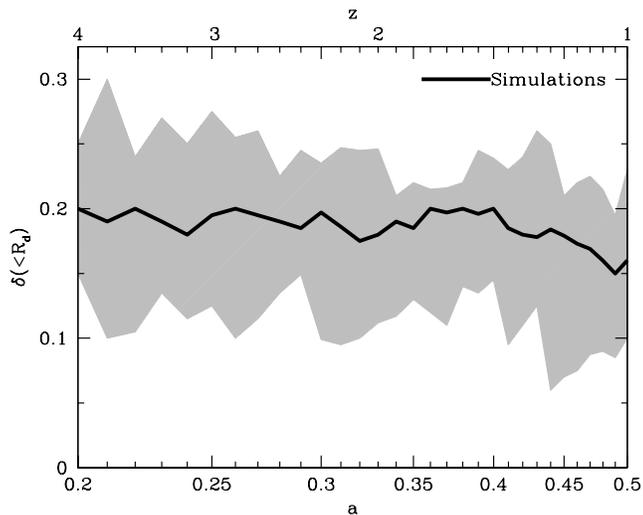}
\caption{
Steady state in disc:
Cold fraction in the disc within $\Rd$, $\delta=M_{\rm cold}/M_{\rm tot}$,
Shown is the median over the simulations (black curve) and the 68 percentiles
(shaded area) for the snapshots where there is net inflow, assumed to undergo
VDI.
The value of $\delta \simeq 0.2$ is consistent with the expectation,
\equ{delta_0.1}, once the mass in cold stars is comparable to the gas mass.
}
\label{fig:delta}
\end{figure}

\smallskip
The cold mass fraction within the disc radius, $\delta$, entering the
right-hand sides of \equ{Mdot_inf} and \equ{Mdot_inf_acc},
is computed for each snapshot inside a sphere of radius $\Rd$.
``Cold" disk stars are defined by $j_z/j_{\rm max}>0.7$,
where $j_z$ is the specific angular momentum of the star particle along the
spin axis of the disc, and $j_{\rm max}=RV_{\rm tot}$
is the maximum specific angular
momentum at the given energy, where $R$ is the radius at the particle position
and $V_{\rm tot}$ is the magnitude of the total particle velocity.
\Fig{delta} shows the average and 68\% scatter of $\delta$ in the simulations.
The median is $\delta \simeq 0.2$ across the whole redshift range, with a 68\%
scatter between 0.12 and 0.25.
The value of $\delta$ due to gas only, $\delta_{\rm g}$,
is about half that value.
These measured values are consistent with the expectations based on
\equ{delta_0.1}.

\smallskip
The dynamical time $\td$ entering the right-hand-side of \equ{Mdot_inf}
is computed for each snapshot from the simulations via \equ{td}.
Based on \equ{Rd}, the parameter $\tlambda$ is derived from 
the radii as determined from the simulations,
and $\tnu = \sqrt{2}$ is assumed.
The Hubble time $t$ is approximated by \equ{a_t}.

\smallskip
For the SFR efficiency in \equ{Mdot_inf_acc} we use $\epsf=0.05$.
This is a high value, estimated to be the effective value of $\epsf$
in our current simulations, dictated
by the requirement that the SFR in the simulations at the given resolution
roughly matches the Kennicutt relation.
This effective value of $\epsf$ is consistent with \equ{sfr}
for the SFR in the simulations
with $\tff=4\Myr$ and where the cold gas available for star formation is
defined to be the mass with temperature lower than $10^4$K and density
higher than $1\cmc$.

\smallskip 
For the analysis here, we have used only snapshots where there is
a net inflow within the disc, as a signature of VDI.
Snapshots where the disc radius is smaller than 3 kpc are excluded; this
minimizes problems related to the spatial resolution of the simulations.
Including these small discs did not make a significant difference at $z<3$,
but did make some difference at $z=3-4$. Since the disc radii at $z>3$ are
uncertain due to wild perturbations, we de-emphasize the results in this
regime even when the disc radius is larger than $3\kpc$.

\smallskip
\Fig{Mdot_inf} compares the specific gas inflow rate in the disc
according to the two sides of \equ{Mdot_inf}. The left-hand side is measured
directly from the simulations, while in the right-hand side we make use
of the $\delta$ and $\td$ as computed from the simulations.
Overall, the toy model is quite successful, at the expected level of
uncertainty.
The average of the left-hand-side is systematically higher than
the right-hand side by a factor of order two in the range $z=3-2$.
\adb{This is at the level of accuracy expected from the crude toy model and the
uncertainties in measuring the simulation results.}
The apparent discrepancy is somewhat larger at $z=4-3$ where the uncertainty
is larger, and somewhat smaller at $z=2-1$.
A better fit could be obtained with either a smaller value of $\gamma$
or a larger value of $\tnu$.

\smallskip
\Fig{Mdot_inf_acc} compares the ratio between the gas inflow rate within the
disc and the gas accretion rate onto the disc as computed by the
two sides of \equ{Mdot_inf_acc}. The values of the parameters used in the
right hand side are $\gamma=1$, $\epsf = 0.05$, and $\fsf=0.5$
The agreement is good except in the redshift range $z=3-4$ where the
uncertainty in disc radius is large.

\section{Steady State in the Disk}
\label{sec:SS_disc}

\subsection{Toy Model: steady state in disc}
\label{sec:SS_disc_toy}

In analogy to the SFR, $\dot{M}_{\rm sf} = \Mg/\tau_{\rm sf}$, based on
\equ{energy} we can express
the inflow rate within the disc as $\dot{M}_{g,in} = \Mg/\tau_{\rm in}$,
where $\tau_{\rm in} = \gamma (\sigma/V)^{-2} \td$.
The gas-mass conservation equation, as in \equ{cont}, 
could then be applied to the disc alone, and include draining by three
processes, namely star formation, outflows, and inflow within
the disc. Based on \equ{sfr} and \equ{energy},  
each of the corresponding sink terms ($i=1,3$) could be expressed in the
form $\dot{M}_{{\rm sink},i}=\Mg/\tau_i$, where $\tau_i = \epsilon_i^{-1}\td$,
so the continuity equation for the disc is
\be
\dot{M}_{\rm g}=\dot{M}_{\rm g,ac}-\dot{M}_{\rm sink} \, ,
\quad \dot{M}_{\rm sink} = \frac{\Mg}{\tau} \, ,
\label{eq:cont_disc}
\ee
where 
\be
\tau^{-1} = \tau_{\rm sf}^{-1} + \tau_{\rm out}^{-1} + \tau_{\rm in}^{-1} \, ,
\label{eq:tau}
\ee
and its steady state solution is analogous to \equ{ss} with $\tau_{\rm sf}$
replaced by the smaller timescale $\tau$.

\smallskip %
\adb{If a steady-state solution exists,}
we can evaluate the steady-state value of $\delta$, the mass fraction of
cold disc inside $\Rd$, following \citet{dsc09}. 
We write 
\be
\delta = \frac{\beta \Md}{\Mb} \, ,
\label{eq:delta_beta}
\ee
where $\beta = \Mb/\Mt(\Rd)$ is the fraction of baryons within the disc
radius, estimated to be $\simeq 0.6$.  
Assuming that $\beta$ is constant, a time derivative of \equ{delta_beta},
combined with \equ{cont_disc}, yields
\be
\dot\delta = \beta(\fg - \beta^{-1}\delta) t_{\rm ac}^{-1} 
- \delta t_{\rm sink}^{-1} (\delta) \, .
\label{eq:dot_delta}
\ee
where $\fg$ is the mass fraction of gas in the baryonic accretion,
and where 
$t_{\rm ac} = \Mb/\dot{\Mb}$ and 
$t_{\rm sink}=\Md/\dot{M}_{\rm sink}$.
Here $\Md$ stands for the cold component in the disc, gas and cold young 
stars.
\adb{If a steady-state solution exists, it will be provided by setting 
$\dot\delta=0$}.
 
\smallskip 
If the sink timescale scales with $\delta^{-2}$ (see below), we write
$t_{\rm sink}/t_{\rm ac} = b \delta^{-2}$, and obtain from \equ{dot_delta} that
the steady-state solution is the solution of the depressed cubic polynomial,
\be
\delta^3 + b\delta -c=0 \, ,
\label{eq:polynomial}
\ee
where $c=b\beta\fg$. 
Its solution is
\be
\delta = u - \frac{b}{3u} , \quad
u=\left( \frac{c}{2} \right)^{1/3}\! 
\left[ 1+\left( 1\!+\!\frac{4b^3}{27c^2} \right)^{1/2} \right]^{1/3}\!.
\label{eq:delta_ss}
\ee
If $b \ll 1$, then $c \ll 1$, and there is no steady-state solution with 
$\delta \sim 1$.
\adb{If $\delta \ll 1$, then there is a solution,
$\delta\simeq c/b = \beta\fg$, which is indeed likely 
to vary rather slowly with time.}
If all three terms of \equ{polynomial} are comparable, with $\delta \sim 0.3$,
then we need to appeal to the full solution, \equ{delta_ss}.
If the gas dominates the cold disc, we can write $t_{\rm sink}=\tau$.
Then the long-term time dependence is via $b \prop a^{-1}$. 
When inserted in \equ{delta_ss}, we find that the value of $\delta$ is expected
to vary rather slowly with time, \adb{consistent with a steady-state solution}.

\begin{figure}
\vskip 7.2cm
\includegraphics{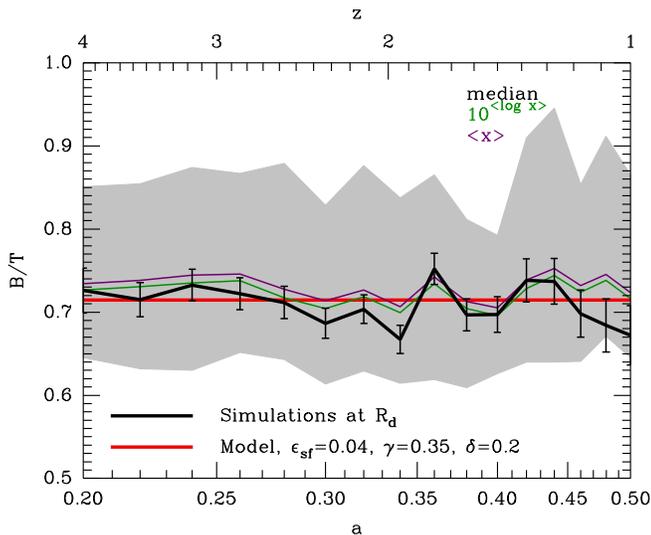}
\caption{
Steady state in disc:
Bulge-to-total ratio within $\Rd$.
\adb{
Shown are the median (thick black), the average (magenta),
and the average of the log within the 90\% percentiles (green)
over the simulated galaxies.
}
Shown in comparison is the toy model prediction,
\equ{Mdot_inf_acc}, with $\delta \simeq 0.2$, $\epsf \simeq 0.04$ and $\fsf
\simeq 0.5$.
The value of B/T and its constancy in time is consistent with the steady-state
bulge growth by VDI, \equ{Mdot_inf_acc},
plus the constancy of $\delta$ as implied by \equ{delta_ss}.
}
\label{fig:BtoT}
\end{figure}

\smallskip 
The scaling of $\tau_{\rm in}$ with $\delta^{-2}$ is apparent from
\equ{Mdot_inf}. The dependence of $\tau_{\rm sf}$ 
(and therefore $\tau_{\rm out}$) 
on $\delta$ could be similar
if star-formation is limited to the disc clumps. 
Then the SFR is proportional to the mass in clumps, namely the
product of clump mass and the number of clumps, $N_{\rm c} \Mc$.
The Toomre clump mass is $\Mc/\Md \simeq (1/4) \delta^2$ \citep{dsc09}. 
Constant values of $N_{\rm c} \Mc$, $N_{\rm c}$ and $\delta$ are consistent
with the slow variation of $\delta$ implied by \equ{delta_ss}.

\subsection{Simulations: steady state in disc}

\Fig{delta} shows that throughout the range $z=3-1$, the average value of 
$\delta$ is 
consistent with being constant, as implied by the steady-state solution
\equ{delta_ss}. A slight decline may be marginally noticeable toward $z \sim 1$.
The average value is $\delta \simeq 0.2$ and the 68\% range is roughly
from 0.12 to 0.24 
It is consistent with the crude estimate in \equ{delta_0.1} once the mass
in cold stars is comparable to the gas mass.

\smallskip
\Fig{BtoT} shows the evolution of bulge-to-total ratio. 
The disc stars are distinguished from bulge stars by $j_z/j_{\rm max}>0.7$.
The average is rather constant at B/T$\simeq 0.73$, with the 68\% range
from $0.63$ to $0.87$. 
This is consistent with the steady-state bulge growth by VDI,
\equ{Mdot_inf_acc}, plus the constancy of $\delta$ as implied by \equ{delta_ss}.
With $\delta \simeq 0.2$, $\epsf \simeq 0.04$ and $\fsf \simeq 0.5$, 
a value of B/T$ \simeq 0.7$ is predicted for $\gamma \simeq 0.35$. 
This is in the ballpark of the value that would yield a good fit in
\fig{Mdot_inf_acc}, testing the validity of \equ{Mdot_inf} for the inflow rate
within the disc.

\section{Conclusion}
\label{sec:conc}

We have verified that key processes of galaxy evolution in its 
most active phase, at $z=1-4$, as captured by hydrodynamical simulations
in a cosmological context, can be \adb{approximated}  
by expressions derived from simple toy models.
\adb{
The systematic evolution of the stacked simulated galaxies is recovered
in the best cases at the level of 10\% accuracy, and in the worse cases 
only to within a factor of two. Even the latter can be useful for qualitative
understanding and back-of-the-envelope estimates.
}
We summarize our main findings below.

\smallskip
\adb{Both the average and the median}
total specific accretion rate into massive haloes are
only weakly dependent on mass and is well approximated as a function of 
redshift by $\propto (1+z)^{5/2}$, \equ{acc_toy}. This includes mergers 
and smoother accretion into the virial radius.  
The systematic growth of virial mass in an individual halo is well 
approximated by a simple exponent in $-z$, \equ{mass_toy}.
This implies a slow variation in time of the absolute accretion rate into a 
given 
halo as it grows in the range $z=5-0.3$, with a maximum near $z\sim 2.2$,
\equ{Mdot_toy}.
\adb{
The distribution of specific accretion rate is lognormal, with a standard
deviation of 0.33 dex.
}
The same expressions describe the accretion of baryons in our simulations.

\smallskip
The penetration of the inflowing baryonic mass ($\Mdot$) 
through the halo into the galaxy 
inside $0.1\Rv$ is $\sim 50\%$ at $z=4-2$, and somewhat larger at $z=2-1$,
\adb{
partly reflecting the higher stellar fraction in the accretion at later times. 
}
On the other hand, the specific inflow rate $\Mdot/M$ 
at $0.1\Rv$ remains very similar to that at $\Rv$,
reflecting the fact that the baryonic mass and its rate of change vary
in a similar way between the two radii. This implies that the
toy-model expressions for $\Mdot/M$ at $\Rv$ 
\adb{can serve as a rather
accurate estimator of the actual characteristic baryon input into 
the central galaxy.
}

\smallskip
The gas in the galaxy obeys a simple mass conservation equation,
\equ{cont}, where the accretion is the source and the star formation 
(and the associated outflows) is the sink. Combined with a universal
star-formation law \citep[e.g.][]{kdm12}, 
the galaxy is predicted to converge to a cosmological 
quasi-steady state, with the SFR following the gas accretion rate into the
central galaxy and the gas mass constant, \equ{ss}.
We note that this may not be true for less massive galaxies and at higher
redshifts, where feedback is likely to be more effective in suppressing star
formation and accumulating a gas reservoir \citep[e.g.][]{zolotov12,kd12}.

\smallskip
The intense inflows into the dense early galaxies trigger violent disc
instability (VDI), which drives mass inflow within the disc, building a 
compact bulge and feeding the central black hole. Toy model estimates
of the inflow rate, performed in several independent ways, 
provide good approximations for the actual inflow rate in the
simulations, both relative to the gas mass and relative to the gas accretion
rate, \equ{Mdot_inf} and \equ{Mdot_inf_acc}. 
A mass conservation equation that is applied to the disc gas alone,
\equ{cont_disc},
predicts a steady state, in which the gas in the disc drains by star 
formation, outflows and inflow within the disc to the bulge, 
and is replenished by cosmological accretion. 
In this steady state, the mass fraction of gas and cold disc stars 
within the disc radius is rather constant in time, and the bulge mass is
comparable to the disc mass or even higher. In the current simulations
$\delta \sim 0.2$ and B/T$\sim 0.6-0.8$, 
while if the gas fraction at $z\sim 2$ is higher, as observed, then $\delta$
could be somewhat larger and B/T somewhat smaller \citep[as in][]{dsc09}. 

\smallskip
We conclude that these toy models can be very useful tools in the study of
key processes, as seen observationally or in simulations,
and can thus help us gain understanding of the complex process of galaxy
formation. 
In turn, the toy models allow us to perform simple analytic calculations 
to predict the expected evolution of galaxies.  
A more detailed treatment of star formation, and the effects of very strong 
outflows, 
are being and will be incorporated into the simulations,
and are yet to be reflected in further testing of toy models.

\section*{Acknowledgments}

We acknowledge stimulating discussions with F.~Bournaud, M.~Krumholz,
N.~Mandelker, E.~Neistein, and R.~Sari.
The simulations were performed at NASA Advanced Supercomputing (NAS) at NASA
Ames Research Center, at the National Energy Research Scientific Computing
Center (NESC) at Lawrence Berkeley Laboratory, and in the astro cluster at
Hebrew University.
This work was supported by ISF grant 24/12,
by GIF grant G-1052-104.7/2009, 
by a DIP grant, 
and by NSF grant AST-1010033.
DC is supported by the JdC subprogramme JCI-2010-07122.



\bibliographystyle{mn2e}
\bibliography{flows}

\appendix
\section{Useful Relations}
\label{sec:useful}

We summarize here the more accurate cosmological relations, valid
in the standard $\Lambda$CDM cosmology at all redshifts, which
the toy-model expressions of this paper approximate in the EdS regime,
$z>1$.
This is largely basic material, based for example on
\citet*{lahav91,carroll92} and \citet{mo02,bd03,db06},
with the addition of accretion rates from \citet{neistein06,neistein08b}.

\subsection{Cosmology}
\label{sec:cosmology}

The basic parameters characterizing a flat cosmological model in the
matter era are the current values of the mean mass density parameter $\omm$
and the Hubble constant $H_0$.
At the time associated with expansion factor $a=1/(1+z)$,
the vacuum-energy density parameter is $\oml(a)=1-\omm(a)$ and
\be
\omm(a)=\frac{\omm\, a^{-3}} {\oml +\omm a^{-3}} \ .
\ee
The Hubble constant is
\be
H(a)=H_0\, (\oml +\omm a^{-3})^{1/2},
\ee
and the age of the universe is
\be
t(a)=\frac{2}{3} H(a)^{-1}\,
\frac{ \sinh^{-1}(|1-\omm(a)|/\omm(a))^{1/2} } {(|1-\omm(a)|)^{1/2}} \ .
\label{eq:tu}
\ee
The mean mass density is
\be
\rho_{u}
\simeq 2.76 \times 10^{-30} {\omm}_{0.3}\, h_{0.7}^2\, a^{-3} \ , 
\label{eq:rhou}
\ee
where ${\omm}_{0.3} \equiv \omm/0.3$,
$h\equiv H_0/100\,{\rm km\,s^{-1}\,Mpc^{-1}}$,
and $h_{0.7}\equiv h/0.7$.

\subsection{Virial relations}
\label{sec:virial}

The virial relations between halo mass, velocity and radius,
\be
\Vv^2 = \frac{G\Mv}{\Rv} ,
\quad \frac{\Mv}{(4\pi/3) \Rv^3} = \Delta \rho_u
\ee
become
\be
V_{200} \simeq 1.02 M_{12}^{1/3} A_{1/3}^{-1/2} \, ,
\quad
R_{100} \simeq 1.03 M_{12}^{1/3} A_{1/3} \, ,
\label{eq:virial}
\ee
where $M_{12}\equiv \Mv/10^{12}\msun$, $V_{200}\equiv \Vv/200\kms$,
      $R_{100} \equiv \Rv/100\kpc$, and
\be
A\equiv (\Delta_{200}\, {\omm}_{0.3}\, h_{0.7}^2)^{-1/3}\, a .
\ee
An approximation for $\Delta(a)$ in a flat universe (Bryan \& Norman 1998) is:
\be
\Delta(a) \simeq (18\pi^2 -82\oml(a)-39\oml(a)^2)/\omm(a) \ .
\label{eq:Delta}
\ee
In the EdS regime, $z>1$, or when referring to $R_{200}$ instead of $\Rv$
at all redshifts, $A \simeq (1+z)^{-1}$ to an accuracy of a few percent.
The virial temperature, defined by
${k\Tv}/{m}=(1/2) \Vv ^2$.
For an isotropic, isothermal sphere, this equals $\sigma^2$, where $\sigma$
is the one-dimensional velocity dispersion and the internal energy
per unit mass is $e=(3/2) \sigma^2$.
Thus
\be
T_6 \simeq 2.87 V_{200}^2 \, ,
\ee
where $T_6 \equiv \Tv/10^6$K.

\subsection{Press Schechter}
\label{sec:PS}

Linear fluctuation growth is given by
\citep{lahav91,carroll92,mo02}
\be
D(a)=\frac{g(a)}{g(1)} \, a ,
\label{eq:da}
\ee
where
\begin{eqnarray}
g(a) 
&\simeq& \frac{5}{2}\omm(a)\\
&\times&
\left[\omm(a)^{4/7}-\oml(a) +\frac{(1+\omm(a)/2)}{(1+\oml(a)/70)} \right]^{-1}
.
\nonumber
\end{eqnarray}

\smallskip
The CDM power spectrum is crudely approximated by
\citep{bbks86}\footnote{Note that in the initial conditions for our simulations
we use the more accurate CAMB numerical solutions based on CAMfast
(Seljak and Zaldarriaga, http://lambda.gsfc.nasa.gov/toolbox/).}
\be
P(k) \propto k\, T^2(k) ,
\ee
with
\begin{eqnarray}
T(k)\!\!\!\!&=&\!\!\!\!\frac{\ln(1+2.34\,q)}{2.34\, q} \\
    \!\!\!\!&\times&\!\!\!\![1 +3.89q +(16.1q)^2 +(5.46q)^3 +(6.71q)^4]^{-1/4}
,
  \nonumber
\end{eqnarray}
where
\be
q=k/(\omm h^2 Mpc^{-1}) .
\ee
It is normalized by $\sigma_8$ at $R=8\hmpc$, where
\be
\sigma^2(R)=\frac{1}{2\pi} \int_0^\infty dk\, k^2\, P(k)\, {\tilde{W}}^2(kR) ,
\ee
and with the Fourier transform of the top-hat window function
\be
\tilde{W}(x)=3(\sin x -x\cos x)/x^3 .
\ee

\smallskip
In the Press Schechter (PS) approximation, the characteristic halo mass
$\Mps(a)$ is defined as the mass of the 1-$\sigma$ fluctuation,
\be
1=\nu(M,a) =\frac{\delta_{\rm c}}{D(a)\, \sigma(M)} ,
\quad \delta_{\rm c}\simeq 1.68,
\label{eq:mstar}
\ee
where $M$ and the comoving radius $R$ are related via the universal density
today: $M=\frac{4\pi}{3} \bar\rho_0\, R^3$.
The mass of 2-$\sigma$ fluctuations is obtained by setting $\nu(M,a)=2$, etc.
Based on the improved formalism of \citet{sheth02},
the fraction of total mass in haloes of masses exceeding $M$ is
\be
F(>M,a) \simeq 0.4 \left( 1 +\frac{0.4}{\nu^{0.4}} \right)
{\rm erfc}\left( \frac{0.85\,\nu}{\sqrt{2}} \right) \ .
\label{eq:sheth}
\ee
This fraction for 1-$\sigma$, 2-$\sigma$, and 3-$\sigma$
fluctuations is 22\%, 4.7\%, and 0.54\% respectively.\footnote{Note that
the Sheth-Tormen approximation becomes quite inaccurate at very high redshifts
\citep{klypin11}.}

\begin{figure}
\vskip 7.2cm
\includegraphics{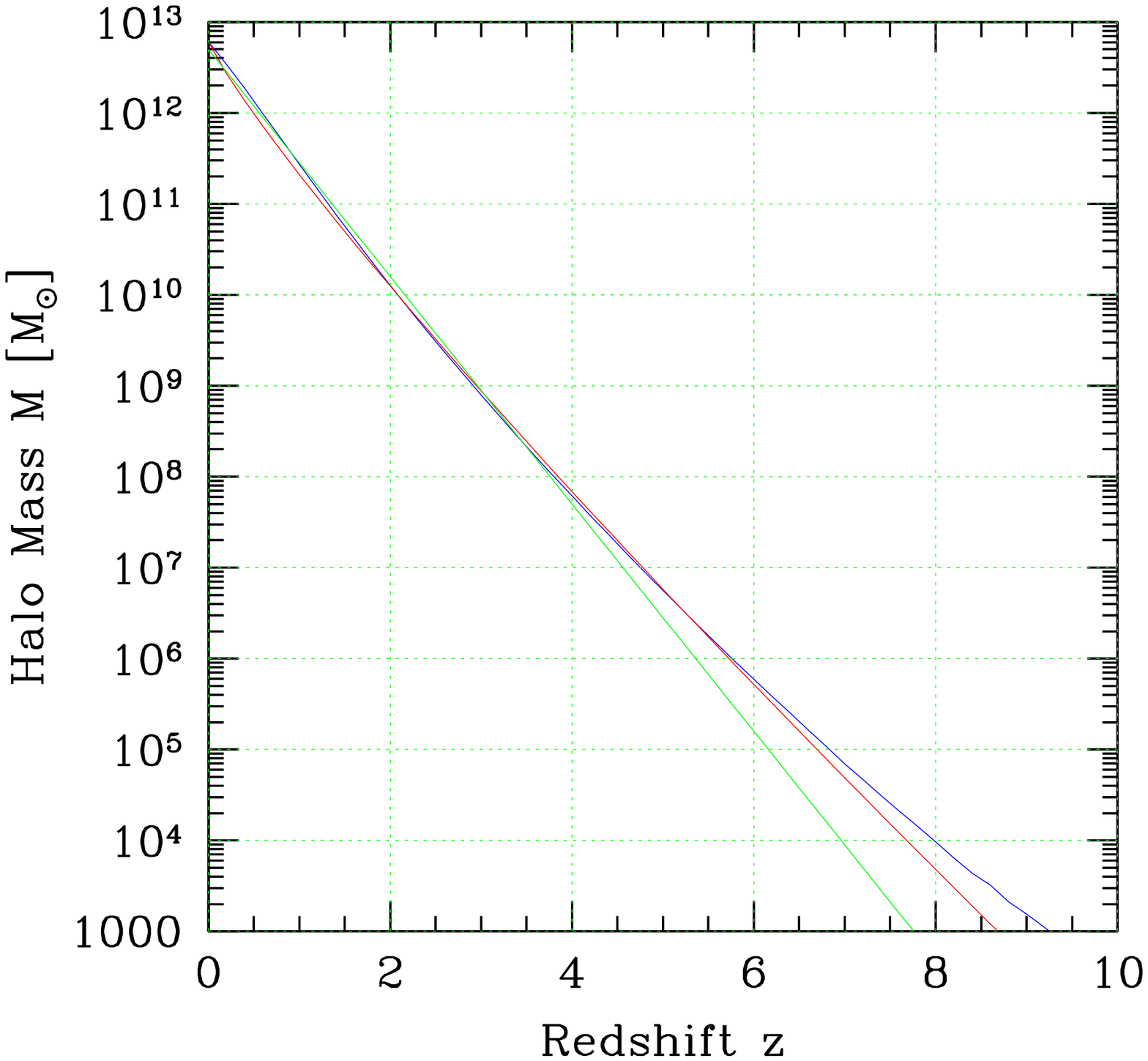}
\caption{The Press-Schechter mass as a function of redshift (blue).
Also shown are the approximations $\log \Mps = 12.7  - 1.29 z$ (green),
and $\log \Mps = 12.78  - 1.46 z^{0.88}$ (red).
}
\label{fig:mstar}
\end{figure}

\smallskip
\Fig{mstar} shows the PS mass $\Mps$ as a function of redshift.
For the standard $\Lambda$CDM with $\omm=0.27$ and $\sigma_8=0.8$
its value at $z=0$ is $M_{\rm ps,0}\simeq 6\times 10^{12}\msun$.
A practical fit for $z \leq 4$ with $\sim 10\%$ accuracy
is provided by a power law in this semi-log plot:
$\log \Mps = 12.7  - 1.29 z$.
At larger redshifts this gradually becomes an underestimate.
A fit that is accurate to $20\%$ in $z<6$ and is an underestimate
by a factor of 2 at $z = 8$ is
$\log \Mps = 12.78  - 1.46 z^{0.88}$.
Trying to provide crude power-law approximations,
we find that $\Mps \prop a^{4.2} \prop t^{3.5}$
are crude approximations in the range $0 \leq z \leq 1$,
and that $\Mps \prop a^{5} \prop t^{4}$ are good to within a factor of 2
in the range $0 \leq z \leq 2$.
These power laws become overestimates at higher redshifts.

\subsection{Accretion Rate}
\label{sec:acc_app}

The average growth of a halo, following its main progenitor, has been derived
by a fit to merger trees from the Millennium cosmological simulation
\citep{springel05}, using a functional form that is motivated by the
EPS approximation \citep{neistein08b}. In terms of the self-similar
dimensionless time variable
$\omega = \delta_{\rm c}[D(a)^{-1}-1]$,
the growth rate of halo of mass $M$ at $\omega$ is
\be
\frac{dM}{d\omega} = -\alpha M^{1+\beta} \, ,
\label{eq:dmdw}
\ee
and the mass at $\omega$, given that the mass at $z=0$ is $M_0$, is
\be
M(\omega|M_0) = (M_0^{-\beta} + \alpha\beta \omega)^{-1/\beta} \, ,
\label{eq:m_w}
\ee
where $M$ is in $10^{12}\msun$.
The value $\beta=0.14$ fits well the Millennium trees for $M$
in the range $10^{11}-10^{14}\msun$, despite the fact that according to EPS
it should vary as $\beta=(n+3)/6$, where $n$ is the local power-spectrum index.
At lower masses, $\beta$ should be smaller, so the mass dependence of the
accretion rate becomes even closer to linear.
For $\sigma_8=0.81$, the best fit is with $\alpha = 0.623$
(it scales as $\prop \sigma_8^{-1}$).
The accuracy of this fit to the Millennium trees is better than 5\% for $z<5$.

\smallskip
The following approximations (better than 0.5\% accuracy at all $z$
for $\omm=0.27$, $h=0.7$, $\sigma_8=0.81$)
relate $\omega$ to $z$,
\be
\omega = 1.28 [(1+z) + 0.086 (1+z)^{-1} +0.22 e^{-1.2 z}] ,
\label{eq:w_z}
\ee
and provide the standard accretion rate $\dot{M}=\dot\omega dM/d\omega$
with
\be
\dot{\omega}= -0.0476 [(1+z) +0.093(1+z)^{-1.22}]^{2.5} \Gyr^{-1} .
\label{eq:wdot}
\ee
The latter scales in proportion to $\prop h_{0.7}$.
Note that in the EdS regime, valid at high $z$, the asymptotic
dependence is $\dot\omega \prop (1+z)^{2.5}$,
because  $D(t) = (1+z)^{-1} \prop t^{2/3}$, so
$\dot{M} \prop t^{-5/3} \prop (1+z)^{5/2}$.

\smallskip
Based on these expressions,
a practical approximation for the typical baryonic accretion rate for
a halo of mass $M$ at $z$ is
\be
\dot{\Mb} = 80\, M_{12}^{1+\beta}\, (1+z)_3^\mu\, f_{0.17} \sy \, ,
\label{eq:acc_app}
\ee
where $f_{0.17}$ is the baryonic fraction in units of 0.17.
With the EdS asymptotic value $\mu=2.5$, the accuracy is $< 5\%$ for $z>1$,
and it becomes an underestimate of $\sim 20\%$ at $z=0$.
With $\mu=2.4$ the accuracy is $<\!5\%$ for $0.2<z<5$,
and is an underestimate of $\sim 10\%$ at $z=0$ and $z=10$.
For most practical purposes we adopt $\beta=0.14$ and $\mu=2.4$.

\begin{figure}
\vskip 7.2cm
\includegraphics{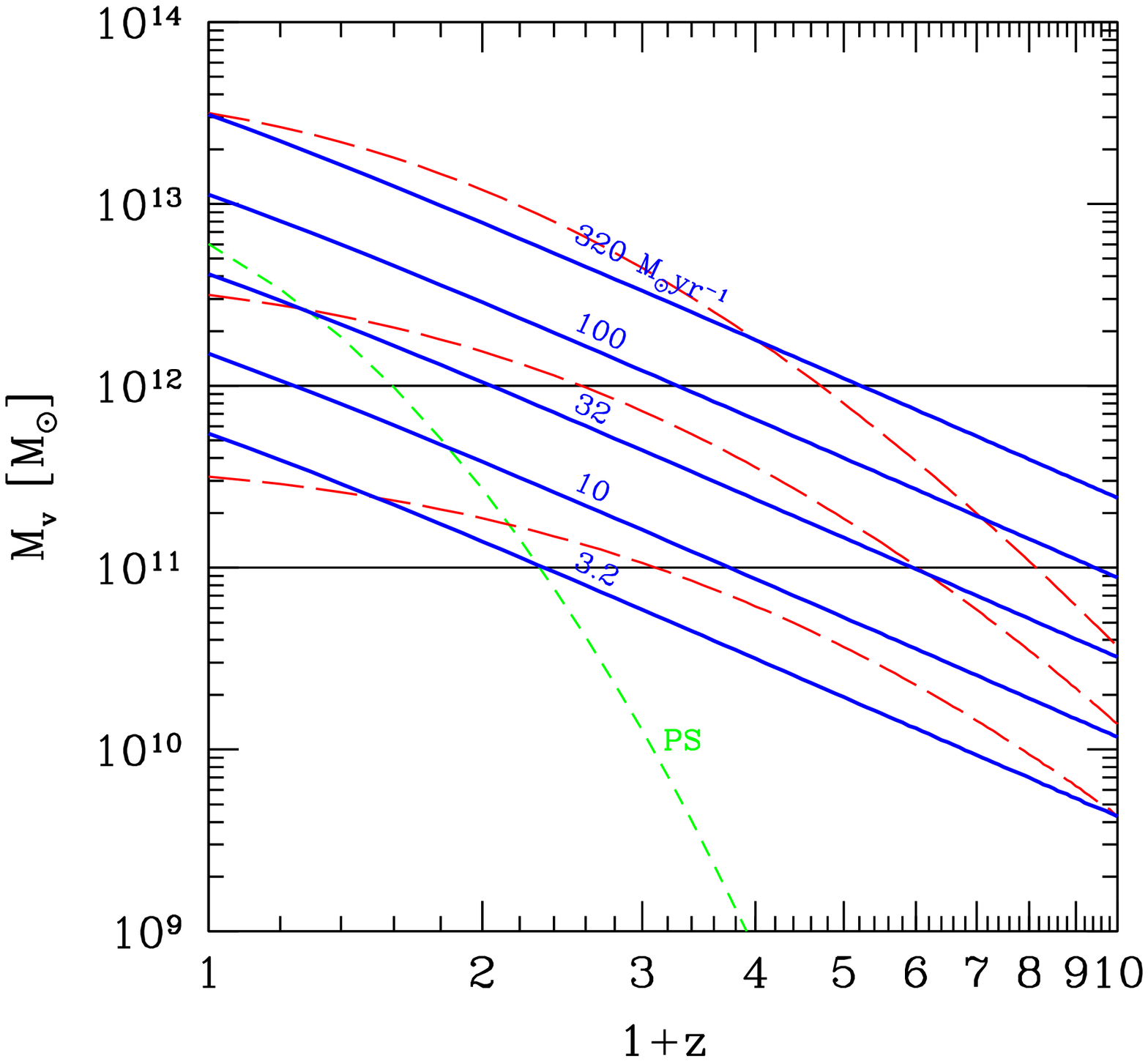}
\caption{Baryonic accretion rate, halo mass and redshift.
The solid blue curves refer to constant values of baryon accretion rate
(an upper limit to the SFR) as marked.
The long-dashed red curves show the average mass growth of the main progenitor
of a halo of a given mass $M_0$ at $z=0$, for $\log M_0 = 11.5, 12.5, 13.5$
from bottom to top.
The short-dash green curve is the Press-Schechter mass.
}
\label{fig:acc}
\end{figure}

\smallskip
\Fig{acc} is a useful summary 
for appreciating the average baryon accretion rate onto
haloes as they grow during cosmological evolution.
Each of the blue curves refers to a given value of $\dot{\Mb}$ (as marked).
They are obtained using the inverse of \equ{dmdw} and \equ{wdot}.
Each of the red curves shows the average mass growth of the main progenitor
of a halo that ends up with a given mass $M_0$ at $z=0$
(for $\log M_0 = 11.5, 12.5, 13.5$), using \equ{m_w} and \equ{w_z}.
For a given halo of $\Mv$ at $z$, one can read the baryon accretion rate
at the corresponding point in the $\Mv,z$ plane by
its position relative to the nearest reference blue curve.
For example, a halo of $\Mv = 10^{12}\msun$ at $z=2$ accretes baryons
at about $80\sy$, as in \equ{acc}.

\smallskip
By following the red curves one can read the history of accretion rate
onto a given halo as it grows.
Take for example the middle red curve, referring to a halo of $M_0=3\times
10^{12}\msun$ at $z=0$. It grew between $z=4$ and $0.5$
from $\Mv \simeq 2\times 10^{11}\msun$ to $2\times 10^{12}\msun$,
while maintaining an accretion rate of $40-60\sy$.

\subsection{Clump Migration by Dynamical Friction}
\label{sec:df}

The timescale for inward clump migration in a marginally unstable disc,
$Q \sim 1$, has been estimated by \citet{dsc09} to roughly be
\be
\tmig \sim \delta^{-2} \td \, ,
\label{eq:tmig_dsc}
\ee
where $\delta$ is the cold mass fraction defined by
$\Md = \delta\, \Mt$, with $\Md$ the cold mass in the disc and 
$\Mt$ the total mass encompassed by the disc radius $\Rd$.
The timescale for evacuating a disc mass $\Md$ is then 
\be
t_{\rm evac} = m_{\rm c}^{-1} \tmig \, , 
\ee
where $m_{\rm c} \sim 0.2$ is the instantaneous fraction of the disc mass in
clumps. This estimate was based on the energy exchange in the
gravitational encounters between clumps.

Here we provide an alternative crude estimate of the clump migration time 
due to the dynamical friction that is exerted on the clump by the off-clump 
disc mass. 
Consider a clump of mass $\Mc$ in a circular orbit with velocity 
$V \simeq (G\Mt/\Rd)^{1/2}$.
The deceleration due to dynamical friction is approximated by the Chandrasekhar
formula \citep[][sec.~8.1]{bt08} to be
\be
\dot V \simeq -4\pi\, G^2\, \Mc\, \rho_{\rm d} \, (\ln\Lambda)\, V^{-2} \, .
\ee
In a $Q \sim 1$ marginally unstable disc, 
the clump mass is related to the disc mass by
\be
\Mc \simeq \delta^2 \Md = \delta^3 \Mt \, ,
\ee
and the disc half thickness $h$ is related to its radius by
\be
h/\Rd \sim \delta 
\ee
\citep[e.g.][]{dsc09}.
Thus, the mean density in the disc is 
\be
\rho_{\rm d} \simeq \frac{\Md}{2\pi\Rd^2 h} \simeq 
\frac{\Mt}{2\pi\Rd^3} \, .
\ee
In the Coulomb logarithm $\Lambda \sim \Md/\Mc \sim \delta^{-2}$ so 
$\ln \Lambda$ is assumed to be $\sim 2$.
We finally obtain
\be
\tmig = V/\dot{V} \sim \frac{1}{4}\,\delta^{-3} \td \, .
\ee
This is indeed comparable to the estimate in \equ{tmig_dsc} for 
$\delta \sim 0.2-0.3$, the typical value in the steady-state solution
of high-$z$ disks \citep{dsc09,cdg12} and in our current simulations, 
\fig{delta}.
The disc evacuation time by clump migration due to dynamical friction is thus 
\be
t_{\rm evac} \sim \delta^{-3}\td \, .
\ee


\label{lastpage}
\end{document}